\documentstyle[preprint,aps]{revtex}
\tightenlines

\newcommand{\ket}[1]{\left|{#1}\right\rangle}
\newcommand{\Hh}{{\textstyle\frac{1}{2}}}
\newcommand{\Dh}{{\textstyle\frac{3}{2}}}
\newcommand{\Fh}{{\textstyle\frac{5}{2}}}
\newcommand{\braket}[2]{\langle{#1}|{#2}\rangle}

\newcommand{\koppl}[3]{ \left[ {#1}\otimes{#2}\right]^{#3} }

\begin{document}
\unitlength1.0cm
\parskip0.5cm

\title{Relativistic Spin-Flavor States in Light Front Dynamics}
\author{M. Beyer and C. Kuhrts}
\address{Fachbereich Physik, Universit\"at Rostock, 18051 Rostock, Germany}
\author{H. J. Weber }
\address{ Institute of Nuclear and Particle Physics, University of
Virginia, \\Charlottesville, VA 22901, USA}
\maketitle
\begin{abstract}
  Orthonormal spin-flavor wave functions of Lorentz covariant quark
  models of the Bakamjian-Thomas type are constructed for nucleon
  resonances. Three different bases are presented. The manifestly
  Lorentz covariant Dirac-Melosh basis is related to the Pauli-Melosh
  basis and the symmetrized Bargmann-Wigner basis that are manifestly
  orthogonal.
\end{abstract}
\par Bakamjian-Thomas-type covariant light cone quark models

\newpage
\section{Introduction}
The light-front form of dynamics introduced by Dirac~\cite{Dir} has by
now become a powerful tool to treat relativistic multi-particle
systems -- as it provides a realization of the Poincar\'e Lie algebra
with a maximal number of kinematical (interaction free) generators. In
particular, the property that certain boosts are free of interactions
is very appealing. For this reason, wave functions of moving frames
may be connected by purely kinematic boosts.  Moreover, with a well
defined Fock expansion, no square root (Hamiltonian) operators and a
simple vacuum structure, light front dynamics represents a viable
framework for relativistic many-body theories. 
\par
As bound systems of three valence quarks predominantly, baryons are
particularly interesting relativistic few-body states. The attractive
features of the front form have motivated many recent calculations of
various form factors of hadrons where rigorous transformation
properties of wave functions under boosts are essential. And, as the
field is rapidly developing, different formulations of relativistic
few-body wave functions have emerged. Particle physicists prefer
constructing multi-quark wave functions of hadrons (or so-called Joffe
currents in the context of QCD sum rule techniques) using Dirac's
gamma matrices. We refer to this approach as the Dirac-Melosh basis in
the following~\cite{Azn,WAP,Dzi,KW,BKK}. Several nuclear physicists
are using Melosh rotated nonrelativistic quark model (NQM) wave
functions~\cite{Ter,KP} which we call the Pauli-Melosh basis below.
For the special case of ground state baryons the spin-isospin
structure of three-quark wave functions can also be rigorously derived
from group theoretical arguments~\cite{Ca}.  At present there is
little communication between these groups, though, which is evident
even from some review articles~\cite{KP,HKT}. Also for this reason we
wish to provide a bridge between these different bases for baryons.
We shall discuss their advantages and disadvantages and construct one
basis from another. Finally, we present one basis that is useful but
hardly known.

Here we do not address any form factor calculations nor related
technical problems such as spurious parts of form factors caused by a
lack of independence from the choice of light cone axis, angular
conditions, $Z$-diagrams, etc. The question as to what dynamical
equations are to be solved is not explored either. Except for the
Pauli-Melosh approach that uses relativistic wave functions that are
generated from a Schr\"odinger equation, this issue is complicated by
the confinement problem of QCD, and no unambiguous solution is known
as yet. The other bases allow one to go beyond the ladder-type
Weinberg equation to include field theoretic effects such as
$Z$-diagrams that are important in form factors and decays, but cannot
be expressed in terms of wave functions alone. Our main goal is to
explain the general relationship between major ingredients and methods
of construction of different bases and clarify their relationships
rather than give a complete review of all states. Such bases are expected 
to play a role in many non-perturbative QCD approaches.  
\par
We start from general kinematics in Sect.II and reconstruct the light
cone momentum variables and Melosh rotations from the by now standard
infinite momentum frame limit.  Since the Pauli-Melosh basis has been
extensively reviewed~\cite{KP} we only give a terse description of it
in Sect.III, while the Dirac-Melosh basis is discussed in greater
detail in Sect.IV, as is the symmetrized Bargmann-Wigner basis in
Sect.V along with its connection to the Dirac-Melosh basis in Sect.VI.

\section{ Kinematics}

It was first shown by Susskind~\cite{Sus} that the infinite momentum
frame (IMF) limit is equivalent to a change of the usual variables
$(t,x,y,z)$ into the light cone variables
$(\tau^+=t+z,x,y,\tau^-=t-z)$.  This is demonstrated in the next
subsection and used to derive the Melosh rotation that transforms the
Dirac spinors into light cone spinors.

\subsection{IMF and Light Cone Variables}

Assume the observer moves with a large negative velocity along the
$z$-axis relative to the baryon rest frame. In the observer's rest
frame the baryon has the energy $E$ and the momentum $P^{\mu
  }=(E,0,0,P)$. A quark has four-momentum $p^{\mu }=(p_0, {\bf
  p}_{\perp},p_z)$, where ${\bf p}_{\perp}=(p_x,p_y)$ and is given
through the boost $L_f(\omega
_p)^{\mu}_{~\nu}\stackrel{\circ}{p}$$^{\nu}$ with
$\stackrel{\circ}{p}$$^{\nu}=(m,0,0,0)$. The quark four-momentum in
the baryon rest frame is denoted by $k^{\mu }$ and obtained by
boosting along the $z$ axis according to $k^{\mu}=L_c^{-1}(\omega
_P)^{\mu}_{~\nu}p^{\nu}$, or explicitly
\begin{eqnarray}\nonumber
k_{0}&=&p_0\cosh \omega -p_z\sinh \omega,
\\\nonumber
k_{z}&=&p_z\cosh \omega -p_0\sinh \omega,
\\
{\bf k}_{\perp}&=&{\bf p}_{\perp},
\label{imf0}
\end{eqnarray}
where $\cosh \omega $ and $\sinh \omega $ are given by
\begin{eqnarray}
\cosh \omega =\frac{E}{M_0},\quad \sinh \omega =\frac{P}{M_0},
\label{imf00}
\end{eqnarray}
and $M_0$ is the free invariant mass of the three quark system,
$M_0^2=P_{\mu }P^{\mu }$, given more explicitly later.  In the
infinite momentum frame (IMF), where $P\rightarrow\infty$, the
longitudinal quark momentum $p_z$ defines the momentum fraction $\eta
$ as
\begin{eqnarray}
p_{z}=\eta P,\quad \lim P\rightarrow\infty.
\label{pz}
\end{eqnarray}
In this limit each quark has a positive $z$-component so that $\eta
>0$ and $\sum_i \eta_i=1$ due to momentum conservation.  We will now
derive the infinite momentum boost given by
$\displaystyle\lim_{P\rightarrow\infty} L_c^{-1}(\omega_{P})L_f(\omega
_p)$.  Expanding the quark energy $p_{0}$ given by
\begin{eqnarray}
p_{0}=\sqrt{\eta ^{2}P^{2}+{\bf p}^{2}_{\perp}+m^{2}},
\label{p0}
\end{eqnarray}
in powers of $1/P$ we obtain in the infinite momentum limit
\begin{eqnarray}
p_{0}=\eta P+\frac{{\bf p}^{2}_{\perp}+m^{2}}{2\eta P}+O(P^{-2}).
\label{imf1}
\end{eqnarray}
Similarly, the baryon energy is expanded as
\begin{eqnarray}
E=P+\frac{M_0^2}{2P}+O(P^{-2}).
\label{imf2}
\end{eqnarray}
Substituting these relations Eqs.~(\ref{imf1}),~(\ref{imf2}) valid for
$P\rightarrow\infty$ into Eq.~(\ref{imf0}) yields finite values for
$k_0$ and $k_z$, the momentum in the nucleon rest frame,
\begin{eqnarray}\nonumber
k_0 &=& \frac{1}{2}\left(\eta M_0+\frac{{\bf p}^2_{\perp}+m^2}
{\eta M_0}\right),\\ \nonumber
k_z &=& \frac{1}{2} \left(\eta M_0 -\frac{{\bf p}^2_{\perp}+m^2}
{\eta M_0}\right),\\
{\bf k}_{\perp }& = &{\bf p}_{\perp }.\label{imf3}
\end{eqnarray}
{}Eq.~(\ref{imf3}) suggests introducing the light cone momentum
components
$$\left(k^{+}=k_{0}+k_{z},\ k^{-}=k_{0}-k_{z},\ {\bf
    k}_{\perp}\right),$$
if we identify the longitudinal fraction
$\eta =p_{z}/P$ with the (kinematically invariant) light cone momentum
fraction $x=p^{+}/P^{+}=k^{+}/M_0$. To see this invariance of $x$,
note that for any pure boost $L_c(\omega)$ in the $z$-direction,
$p^{+}$ and $p^{-}$ scale according to
\begin{eqnarray}\nonumber
L_c(\omega)p^{\pm}&=&L_c(\omega)\left(p_0\pm p_z\right)\\
&=&\left(\cosh\omega \pm\sinh\omega \right)p^{\pm}=
\exp\left(\pm\omega \right)p^{\pm}.
\label{eqn:zboost}
\end{eqnarray}
Thus, Eq.~(\ref{imf3}) becomes a simple
transformation of momentum variables~\cite{Sus,Kog}
\begin{eqnarray}\nonumber
k_{0}&=&\frac{1}{2}\left(x M_0+\frac{{\bf k}^{2}_{\perp}
+m^{2}}{x M_0}\right)=\frac{1}{2}\left(k^{+}+k^{-}\right),\\\nonumber
k_{z}&=&\frac{1}{2} \left(x M_0-\frac{{\bf k}^{2}_{\perp}+m^{2}}{xM_0}\right)
=\frac{1}{2}\left(k^{+}-k^{-}\right),\\
{\bf k}_{\perp}&=&{\bf p}_{\perp},
\label{imf4}
\end{eqnarray}
since the invariant quark mass $m$ and the light-cone energy variable
$k^-$ are related by
\begin{eqnarray}
m^{2}=k^{+}k^{-}-{\bf k}^{2}_{\perp},\quad
k^{-}&=&\frac{{\bf k}^{2}_{\perp}+m^{2}}{k^{+}}.
\end{eqnarray}
The scalar product for light cone coordinates is
\begin{eqnarray}
a\cdot b=a_{\mu}b^{\mu}=\frac{1}{2}\left(a^{+}b^{-}+a^{-}b^{+}\right)
-{\bf a}_{\perp}{\bf b}_{\perp},
\end{eqnarray}
For later purpose we will generalize Eq.~(\ref{eqn:zboost}) to
arbitrary boost directions {\boldmath{$\omega$}}. For a moving quark
with momentum $p^{\mu}$ the transformed components are given by
\begin{eqnarray}\nonumber
p'^+&=&p^+\exp{(\omega _z)},\\
\mbox{\boldmath{$p$}}'_{\perp}&=
&\mbox{\boldmath{$p$}}_{\perp}+
\mbox{\boldmath{$\omega$}} _{\perp}p^+\exp{(\omega _z)}.
\label{gboo}
\end{eqnarray}

{}Eq.~(\ref{imf4}) is the main result of this paragraph. It defines a
boost from the rest system of a quark to the rest system of the
nucleon given in light cone coordinates, viz.  $k^\mu=
L_{cf}(\omega_{k})^{\mu}_{~\nu}{\stackrel{\circ}{p}}$$^{\nu}$. To find
the proper transformation of the spinors we need the $SL(2,C)$
representation of $L_{cf}$ which will be derived in the
next subsection.

\subsection{The Melosh Rotation}\label{Melsub}
The  proper transformation of the instant form spin states
$ |\vec{k} s^{\prime}\rangle_{{\rm inst}} $
induced by the transformation to the light cone
is given by
\begin{eqnarray}\nonumber
|k^{+},{\bf k}_{\perp} s\rangle&=&\sum
  R_{s^{\prime}s}\: |\vec{k},s^{\prime}\rangle_{\rm inst},\\
|\vec{k} s\rangle_{\rm inst}&=&\sum R^{-1}_{s^{\prime}s}\:|k^{+},
{\bf k}_{\perp} s^{\prime} \rangle,
\label{melrot}
\end{eqnarray}
where $R_{s^{\prime}s}$ is the Melosh rotation.  We now derive the explicit
form of the Melosh rotation matrix for Dirac spinors.
To do so, we use the
above demonstrated equivalence between the light cone frame and the
infinite momentum frame. We will summarize this subsection in terms of a
more elegant $SL(2,C)$ approach.
\par
The instant-form Dirac spinor is
written in the representation given, e. g., in Ref.~\cite{BjD},
\begin{eqnarray}
u(\vec{p},s)=\sqrt{\frac{p_{0}+m}{2m}}
\left(\begin{array}{c}\chi(s)\\
\frac{\vec{\sigma}\cdot
\vec{p}}{p_{0}+m}\chi(s)\end{array}\right),
\end{eqnarray}
from which the light cone spinor $u_{LC}$ is obtained by a boost to the IMF
(where $p$ depends on $P$, see Eqs.~(\ref{pz}),(\ref{p0})) and
replacing $\eta \rightarrow x$ as explained in the previous subsection
\begin{eqnarray}
u_{LC}\left(k^{+},{\bf k}_{\perp},s\right)&=
&\lim_{P\rightarrow \infty} u(\vec{p},s),
\end{eqnarray}
with the spinor transformation matrix
(corresponding to the boost in the negative z-direction)
\begin{eqnarray}
S\left(k\leftarrow p\right)=\cosh\frac{\omega}{2}
-\alpha_{3}\sinh\frac{\omega}{2}.
\end{eqnarray}
The explicit form of $u_{LC}$ then is
\begin{eqnarray}\nonumber
u_{LC}\left(k^{+},{\bf k}_{\perp},s\right) =\lim_{P\rightarrow\infty}
\frac{1}{\sqrt{2m\left(p_{0}+m\right)}}
\left(\begin{array}{c}
\left[\cosh\frac{\omega}{2}\:\left(p_{0}+m\right)
-\sinh\frac{\omega}{2}\:\sigma_{3}\vec{\sigma}\cdot
\vec{p}\right]\chi(s)\\ \\
\left[\cosh\frac{\omega}{2}
\:\vec{\sigma}\cdot \vec{p}-\sinh\frac{\omega}{2}\:
\left(p_{0}+m\right)\sigma_{3} \right]\chi
(s)\end{array}\right),
\end{eqnarray}
upon using Eq.~(\ref{imf00}),
\begin{eqnarray}\nonumber
\cosh\frac{\omega}{2}&=&\sqrt{\frac{E+M_0}{2M_0}},\quad
\sinh\frac{\omega}{2}=\sqrt{\frac{E+M_0}{2M_0}}\frac{P}{E+M_0}\simeq
\sqrt{\frac{E+M_0}{2M_0}}\left(1-\frac{M_0}{P}\right)
\end{eqnarray}
\begin{eqnarray}
p_{0} = \eta P+\frac{{\bf p}^{2}_{\perp}+m^{2}}{2\eta P}.
\end{eqnarray}
The last approximation holds neglecting $O(P^{-2})$.
Hence we obtain for $u_{LC}$
\begin{eqnarray}\nonumber
u_{LC}\left(k^{+},{\bf k}_{\perp},s\right) &=&\lim_{P\rightarrow\infty}
\frac{1}{\sqrt{2m\left(p_{0}
+m\right)}}\sqrt{\frac{E+M_0}{2M_0}}
\end{eqnarray}
\begin{eqnarray}\nonumber
&\times&\left(\begin{array}{c}\left[\eta P+\frac{{\bf
  p}^{2}_{\perp}+m^{2}}{2\eta P}+m-\left(-\mbox{\boldmath$\sigma$}_{\perp}
\cdot {\bf p}_{\perp}\sigma_{3}+\eta
P\right)\left(1-\frac{M_0}{P}\right)\right]\chi(s)\\ \\
\left[\mbox{\boldmath$\sigma$}_{\perp}\cdot {\bf p}_{\perp}
+\eta P\sigma_{3}-\left(\eta P+\frac{{\bf p}^{2}_{\perp}+m^{2}}
{2\eta P}+m\right)\sigma_{3}
\left(1-\frac{M_0}{P}\right)\right]\chi (s)
\end{array}\right) .\\
\end{eqnarray}
Here the leading orders in $P$ cancel and the remaining expression
is independent of the momentum $P$, viz.
\begin{eqnarray}\nonumber
u_{LC}\left(k^{+},{\bf k}_{\perp},s\right)
&=&\frac{1}{2\sqrt{m\eta M_0}}\left(\begin{array}{c}\left[m+\eta M_0+
\mbox{\boldmath$\sigma$}_{\perp}\cdot {\bf k}_{\perp}\sigma_{3}\right]
\chi(s)\\ \\
\left[\mbox{\boldmath$\sigma$}_{\perp}\cdot {\bf k}_{\perp}-m\sigma_{3}+\eta
  M_0\right]\chi(s)\end{array}\right)\\
&=&\frac{1}{2\sqrt{mk^{+}}}\left(\begin{array}{c}\left(k^{+}+m\right)\chi(s)+
\mbox{\boldmath$\sigma$}_{\perp}\cdot {\bf k}_{\perp}\sigma_{3}\chi(s)\\ \\
\left(k^{+}-m\right)\sigma_{3}\chi(s)+\mbox{\boldmath$\sigma$}_{\perp}\cdot
{\bf k}_{\perp}\chi(s)
\end{array}\right).
\label{pmelosh}
\end{eqnarray}
Introducing the standard form of the Pauli spinors~\cite{BjD} finally leads to
\begin{eqnarray}\nonumber
u_{LC}\left(k^{+},{\bf k}_{\perp},\uparrow\right)&=&
\frac{1}{2\sqrt{mk^{+}}}\left(\begin{array}{c}k^{+}+m\\k^{R}\\k^{+}-m\\k^{R}
\end{array}\right),\\
u_{LC}\left(k^{+},{\bf k}_{\perp},\downarrow\right)&=&
\frac{1}{2\sqrt{mk^{+}}}\left(\begin{array}{c}-k^{L}\\k^{+}+m\\k^{L}\\-k^{+}+m
\end{array}\right),
\end{eqnarray}
where $k^{L,R}=k_{x}\mp ik_{y}$.
Similarly for the $v$-spinors
\begin{eqnarray}\nonumber
v_{LC}\left(k^{+},{\bf k}_{\perp},\uparrow\right)&=&
\frac{1}{2\sqrt{mk^{+}}}\left(\begin{array}{c}k^{+}-m\\k^{R}\\k^{+}+m\\k^{R}
\end{array}\right),\\
v_{LC}\left(k^{+},{\bf k}_{\perp},\downarrow\right)&=&
\frac{1}{2\sqrt{mk^{+}}}\left(\begin{array}{c}k^{L}\\-k^{+}+m\\-k^{L}\\k^{+}+m
\end{array}\right).
\label{mela}
\end{eqnarray}
For the Dirac spinor the Melosh rotation $R_{s^{\prime}s}$
defined in Eq.~(\ref{melrot}) is thus given by
\begin{eqnarray}\nonumber
u\left(\vec{k},\uparrow\right)&=&\frac{1}{\sqrt{2k^{+}\left(k_{0}+m\right)}}
\left[\left(k^{+}+m\right)u_{LC}\left(k^{+},
{\bf k}_{\perp},\uparrow\right)-k^{R}u_{LC}\left(k^{+},{\bf k}_{\perp},
\downarrow\right)\right],\\
u\left(\vec{k},\downarrow\right)&=&\frac{1}{\sqrt{2k^{+}
\left(k_{0}+m\right)}}\left[\left(k^{+}+m\right)u_{LC}\left(k^{+},
{\bf k}_{\perp},\downarrow\right)+k^{L}u_{LC}\left(k^{+},{\bf
  k}_{\perp},\uparrow\right)\right].\hspace{1 cm}
\label{dirmel}
\end{eqnarray}
If we combine spin $\uparrow$ and spin $\downarrow$ into the
fundamental representation of $SU(2)$, the matrix representation of $R$ is
given as
\begin{equation}
\left(\uparrow,\downarrow\right)_{\rm inst}=
\left(\uparrow,\downarrow\right)_{\rm LC}
\left(\begin{array}{cc}k^++m & k^L\\ -k^R & k^++m
\end{array}\right) {1\over \sqrt{2k^{+}(m+k_0)}},
\label{Dmelosh}
\end{equation}
or
\begin{equation}
R={(k^{+}+m)+\mbox{\boldmath$\sigma $}_{\perp}\cdot {\bf k}_{\perp}
\sigma _{3}\over \sqrt{2k^{+}(m+k_0)}},
\end{equation}
which coincides with Refs.~\cite{Ter,KP}.
\par
Alternatively, we now calculate the light cone spinor
$u_{LC}(k^+,{\bf k}_{\perp},s)$ given in Eq.~(\ref{pmelosh}) using the well
known relation of proper Lorentz transformations to the $SL(2,C)$
covering group (see e.g.~\cite{thaller}). The respective $SL(2,C)$
representation of the IMF boost, Eq.~(\ref{imf4}), is given by
\begin{equation}
\sigma_\mu k^\mu = A\,  \sigma_\mu{\stackrel{\circ}{p}}
\mbox{$^\mu$}\, A^\dagger,
\end{equation}
where $\sigma^\mu=(1,\sigma^i)$. The matrix $A\in SL(2,C)$
corresponding to the transformation given in Eq.~(\ref{imf4}) takes the form
\begin{equation}
A=\frac{1}{\sqrt{k^+m}}
\left(\begin{array}{cc} k^+ &0\\k_R&m \end{array}\right).
\label{ALC}
\end{equation}
The corresponding transformation of Dirac spinors to the light cone that is
required to give a complete representation of the
full Lorentz group is given by~\cite{thaller}
\begin{equation}
S_{LC}=U\, \left(\begin{array}{cc}
A&0\\0&(A^\dagger)^{-1}
\end{array}\right)\, U^\dagger,\qquad
U=\frac{1}{\sqrt{2}}\left(\begin{array}{cc}
1&1\\1&-1\end{array}\right),
\end{equation}
where $U$ transforms to the diagonal Weyl representation of $S_{LC}$.
The explicit form using Eq.~(\ref{ALC}) is then given by
\begin{equation}
S_{LC}(k)=
\frac{1}{2\sqrt{mk^{+}}}\left(\begin{array}{ll}
\left(k^{+}+m\right)+\mbox{\boldmath$\sigma$}_{\perp}\cdot {\bf k}_{\perp}
\sigma_{3}
&~\left(k^{+}-m\right)\sigma_{3}+\mbox{\boldmath$\sigma$}_{\perp}\cdot
{\bf k}_{\perp}\\
\left(k^{+}-m\right)\sigma_{3}+\mbox{\boldmath$\sigma$}_{\perp}\cdot
{\bf k}_{\perp}
&~\left(k^{+}+m\right)+\mbox{\boldmath$\sigma$}_{\perp}\cdot {\bf k}_{\perp}
\sigma_{3}
\end{array}\right).
\label{eqn:Dirmel}
\end{equation}
Using $u_{LC}(k^+,{\bf k}_{\perp},s)=S_{LC}(k) u(\vec{0},s)$ then leads to
Eq.~(\ref{pmelosh}).

\subsection{Three-quark coordinates}
The quark momentum variables for baryons in an arbitrary frame are
denoted by $p_{1},$ $p_{2},$ $p_{3}$, while those in the baryon rest
frame are written as $k_1,$ $k_2,$ $k_3$, with $\sum_{i=1}^3 \vec{k}_i
=0$.  The relation between $p_i$ and $k_i$ are given by
Eqs.~(\ref{gboo}), (\ref{imf4})
\begin{eqnarray}\nonumber
{\bf k}_{i,\perp}&=& {\bf p}_{i,\perp}-x_{i}{\bf P}_{\perp},\quad
x_{i}=\frac{{p}_{i}^{+}}{{P}^{+}}\\\nonumber
k_{i,z}&=&k^{+}-k_{0} = x_{i}M_0-\frac{P\cdot p}{M_0}\\
&=&x_{i}M_0-\left(\frac{m_{i}^{2}+\left({\bf p}_{i\perp}-x_{i}
{\bf P}_{\perp}\right)^{2}}{2x_{i}M_0}+\frac{x_{i}M_0}{2}\right),
\label{kin}
\end{eqnarray}
where $P$ is the total free four-momentum of the three-quark system and
$M_{0}$ the invariant mass squared
\begin{eqnarray}
M_{0}^2=\sum_i {{m_i^2+{\bf k}_{i,\perp}^2}\over x_i}.\label{invmass}
\end{eqnarray}
For proper symmetrization in the quark indices and equal quark masses
$m_i=m_q$ we adopt the standard normalized Lovelace coordinates
\begin{eqnarray}\nonumber
\vec{k}_{\rho }&=&\frac{1}{\sqrt{2}}\left(\vec{k}_{1}-\vec{k}_{2}\right)
,\\
\vec{k}_{\lambda }&=&\frac{1}{\sqrt{6}}\left(\vec{k}_{1}+\vec{k}_{2}
-2\vec{k}_{3}\right).
\label{rholam}
\end{eqnarray}
Often used relativistic
alternatives are the relative 4-momentum (space-like Jacobi) variables
\begin{eqnarray}
q_3={x_2p_1-x_1p_2\over x_1+x_2},\ Q_3=(1-x_3)p_3-x_3(p_1+p_2),
\label{rmom}
\end{eqnarray}
for the $\perp$ and + components,
and $q_1, Q_1, q_2, Q_2$ from cyclic permutation of the indices. Note that
${\bf Q}_{i\perp}={\bf k}_{i\perp}$ for i=1,2,3.

\section{Pauli-Melosh basis}
We now present the three quark basis generated by rotating the
nonrelativistic wave function to the light cone~\cite{KP}.  Single
particle state vectors $|p^{+}, {\bf p}_{\perp} \lambda \rangle$ are
labeled by the momentum $(p^{+}, {\bf p}_{\perp})$ and spin projection
$\lambda$ written in light front coordinates so that the mass shell
condition $p^- = (m^2+{\bf p}_{\perp}^2)/p^+$ is satisfied. Under a
light front boost $p'^{\mu }=L ^{\mu }_{~\nu }\,p^{\nu }$, the state
vectors transform unitarily as $U(L )|p^{+}, {\bf p}_{\perp} \lambda
\rangle=\sqrt{{p'^+/ p^+}}\;|p^{'+}, {\bf p} ^{\prime}_{\perp} \lambda
\rangle$, where the light front spin $\lambda $ remains unchanged (no
Wigner rotation). These states are related to those of the instant
form, $|\vec{p} m_s \rangle$, by the Melosh rotation $R_{cf} =
\underline{L}^{-1}_c(\vec{p}) \underline{L}_f(\vec{p})$ (where
$\underline{L}$ denotes the $SL(2,C)$ representation of $L$) so that
$$|p^{+}, {\bf p}_{\perp} \lambda \rangle =\sum_{m_s}\sqrt{E(\vec{p})/
  p^+}|\vec{p} m_s \rangle D^{1/2}_{m_s \lambda}(R_{cf}),$$
which corresponds to Eq.~(\ref{melrot}).  Baryon three-quark states
$|j;\vec{P}
\lambda \rangle$ with spin $j$, spin projection $\lambda$ and momentum
$\vec{P}$ are related to wave functions according to
\begin{eqnarray}\nonumber
\langle \vec{p}_1 \lambda _1 \vec{p}_2 \lambda _2 \vec{p}_3 \lambda
_3|j;
\vec{P}
\lambda \rangle&=&\left|{\partial\left(\vec{p}_1, \vec{p}_2,
    \vec{p}_3\right)\over
\partial
\left(\vec{P}, \vec{k}_{\rho }, \vec{k}_{\lambda }\right)}\right|^{-1/2}
(2\pi )^3\delta (\sum_i \vec{p}_i-\vec{P})\\
&&\times \sum_{m's}\langle {1\over 2}m_1 {1\over 2}m_2|s_{12}m_{12}\rangle
\langle s_{12}m_{12}{1\over 2}m_3|sm_s\rangle
\langle l_{\rho }m_{\rho }l_{\lambda }m_{\lambda }|Lm_L\rangle\\
\nonumber
&&\times \langle Lm_Lsm_s|jm\rangle
Y_{l_{\rho }m_{\rho}}\left(\hat{\bf k}_{\rho }\right)
Y_{l_{\lambda }m_{\lambda }}\left(\hat{\bf k}_{\lambda }\right)
\Phi(k_{\lambda},k_{\rho})\\
&&\times D^{1/2 \dagger}_{m_1 \lambda _1}(R(k_1))
D^{1/2 \dagger}_{m_2 \lambda _1}(R(k_2))
D^{1/2 \dagger}_{m_3 \lambda _1}(R(k_3)),
\label{P-M}
\end{eqnarray}
with obvious notation for the $SU(2)$ Clebsch-Gordan coefficients.  The
Jacobian is $$\left|{\partial\left(\vec{p}_1, \vec{p}_2,
      \vec{p}_3\right) \over \partial \left(\vec{P}, \vec{k}_{\rho },
      \vec{k}_{\lambda }\right)}\right| ={p_1^+p_2^+p_3^+M_0\over
  E(\vec{k}_1)E(\vec{k}_2)E(\vec{k}_3)P^+}.$$
The totally symmetric
momentum wave functions of the bound state, which are not shown above,
are separately orthogonal.  Because of the orthogonality of the Melosh
rotations, this basis of wave functions is manifestly orthogonal,
which is clearly an advantage of the Pauli-Melosh basis. In contrast,
Lorentz invariance is not manifest. Note, however, that relativistic
models of the Bakamjian-Thomas type, where an interaction that
commutes with the total spin is added to $M_0$, are Lorentz invariant
if their interactions are rotationally invariant in terms of 3-vector
(momentum or coordinate, etc.) variables of the
particles~\cite{Ter,KP,BT}. This method of dealing with an 
interaction differs from that of field-theoretic Lagrangians and affects 
the connection with Feynman (and light-cone time-ordered) diagrams. A 
closely related light-front field theory~\cite{MK} maintains the 
connection with light-cone time-ordered diagrams with interactions that 
do not necessarily commute with the total spin; it has different off-shell 
properties. A Dirac-Melosh basis is also constructed there that includes some 
states from configuration mixing but not all. Despite the three-vector 
appearance in Eq.~(\ref{P-M}), the quark momentum variables are relativistic, 
their $z$-components being defined in Eq.~(\ref{imf4}) in terms of their
longitudinal momentum fraction $x_i$ of Eq.~(\ref{kin}) and $M_0$ of
Eq.~(\ref{invmass}). The quark light-cone Pauli spinors depend on the
relativistic quark momentum variables via the Melosh transformation.
Therefore, despite working only with Pauli spinors (and coupling them
by $SU(2)$ Clebsch-Gordan coefficients as in Eq.~(\ref{P-M})), the
Pauli-Melosh basis provides a consistent relativistic many-body
framework for baryons. Small Dirac components are not necessary in
such theories. In this sense, then, the Pauli-Melosh states form a
minimal relativistic basis that is viable as long as pre-existing (or
intrinsic) quark-antiquark excitations in baryons may be safely
neglected at the low-energy scale $\Lambda_{QCD}$. As we shall see
below, the Dirac-Melosh basis contains many-body states with small
Dirac components in addition to the Pauli-Melosh basis.

In the baryon rest frame, the
Pauli-Melosh basis becomes the usual NQM basis in the nonrelativistic
limit, where the Melosh rotation is replaced by unity.  Of course,
this basis is far from complete as a relativistic Fock state basis.
Only kinematic relativistic effects are included and boosts in
particular. Dynamic (or intrinsic) quark-antiquark Fock components are
not included but can be added in terms of the basis states of the
Dirac-Melosh type shown for the nucleon in Table~\ref{tab:1} and for
nucleon resonances in Tables~\ref{tab:2} and ~\ref{tab:3} in the next
Sect. IV. 

\section{Baryon spin-isospin states in the Dirac-Melosh basis}

The NQM basis of hadron wave functions can be mapped one-to-one onto
relativistic multi-quark light cone states. The first step in the
construction of the Dirac-Melosh basis from the NQM basis consists in
transforming the Clebsch-Gordan coefficients of the $SU(2)$ group into
products of spin-isospin (flavor) matrix elements between quark and
total momentum light cone spinors using the Wigner-Eckart theorem.
Thereby, momentum dependence enters into the relativistic spin-flavor
wave functions which, in the NQM, do not depend on momentum, thus
removing manifest orthogonality from the Dirac-Melosh basis. A few
examples of nucleon resonance states will serve us to illustrate this
part of our procedure, e.g., the $N(938)$ is introduced in the next
subsection, and subsequently the $N^*_{\frac{1}{2}^{-}}(1535),
N^*_{\frac{3}{2}^{-}}(1520), N^*_{\frac{5}{2}^{-}}(1675)$.

\subsection{The nucleon spin-isospin states}
The Wigner-Eckart theorem allows one to rewrite the Clebsch-Gordan
coefficients in the nonrelativistic nucleon spin-isospin wave function
$|(s_{12}\frac{1}{2})J=\frac{1}{2}\ M_{J}=\lambda\rangle\times
|(t_{12}\frac{1}{2})T=\frac{1}{2}\ M_{T}\rangle $ in terms of two
products between Pauli spinors~\cite{Azn,WAP,Dzi}
\begin{eqnarray}\nonumber
|(0\Hh)\Hh\lambda\rangle &=&
\sqrt{\frac{1}{2}}\sum_{m_{1}m_{2}m_{3}}\left(\chi_{m_{1}}^{\dagger}i
\sigma_{2}\chi_{m_{2}}^{\ast}\right)\left(\chi_{m_{3}}^{\dagger}
\chi_{\lambda}\right)\chi_{m_{1}}\chi_{m_{2}}\chi_{m_{3}},\\
|(1\Hh){\Hh}\lambda\rangle &=&
-\sqrt{\frac{1}{6}}\sum_{m_{1}m_{2}m_{3}}\left(\chi_{m_{1}}^{\dagger}
\vec{\sigma}i\sigma_{2}\chi_{m_{2}}^{\ast}\right)\cdot\left(
\chi_{m_{3}}^{\dagger}\vec{\sigma}
\chi_{\lambda}\right)\chi_{m_{1}}\chi_{m_{2}}\chi_{m_{3}},
\label{nspin}
\end{eqnarray}
The associated isospin matrix elements have the same form as
Eq.~(\ref{nspin}). Under a Melosh rotation of the Pauli spinors,
Eq.~(\ref{melrot}), to the light cone, Eqs.~(\ref{nspin}) retain their
form. After introducing light cone spinors this equation then matches
the spin part of the nucleon wave function of the Pauli-Melosh basis
given in the previous section. The generalization of
Eqs.~(\ref{nspin}) to covariant expressions will now be
achieved in two further steps.  First, we express the Pauli light cone
spinors by the Dirac light cone spinors given in the nucleon rest
system. In a second step this will be generalized to arbitrary nucleon
momenta.

The light cone Pauli spinor is given by the upper part of the light
cone Dirac spinor (just as it is the case for the instant form Dirac
spinor). This may be formally written as a projection, viz.
\begin{equation}\chi_{LC}=N\;[1,0]\;u_{LC}(k),
\end{equation}
where $[1,0]$ is understood as a $2\times 4$ matrix written in block
notation. Also a normalization factor $N= \sqrt{2m/(k_0+m)}$ has been
introduced to respect proper normalization. Therefore, we are able to
replace the matrix products between Pauli spinors by products between
light cone Dirac spinors. As an illustration we write
\begin{equation}
\chi_{LC,m_1}^{\dagger}i\sigma_{2}\chi_{LC,m_2}^{\ast}=
N_1N_2 \;\bar{u}_{LC}(k_1)\,[1,0]^{\top}i\sigma_{2}[1,0]\,
\bar{u}_{LC}^{\top}(k_2),
\label{firststepA}
\end{equation}
where $N_1=N(k_1)$. Expressing the matrix element in
terms of standard Dirac matrices leads to
\begin{equation}
[1,0]^{\top}i\sigma_{2}[1,0]= \frac{1}{2}
(1+\gamma _0)\gamma _5 i\gamma _0\gamma _2
(1+\gamma _0).
\label{firststepB}
\end{equation}
The other matrix elements are rewritten in a similar way.

To write the resulting expressions in an arbitrary frame moving with
the nucleon momentum $(P^+,{\bf P}_{\perp})$, i.e. for each quark
$p^{\mu}=L(\omega_{P}){^\mu}_{\nu}k^{\nu}$, note that each quark
spinor experiences a transformation according to
\begin{equation}
\bar u_{LC}(k)=\bar u_{LC}(p) \,\bar S(\omega_{P}).
\label{midA}
\end{equation}
The $4\times 4$ matrix $\bar S(\omega_{P})=\gamma_{0}
S^{+}(\omega_{P})\gamma_{0}$ may be written in the following way
(compare Ref.~\cite{BjD} for instant form spinors)
\begin{equation}
\bar S(w_P)=\left[u_{LC}^{\uparrow}(P),u_{LC}
^{\downarrow}(P),v_{LC}^{\uparrow}(P),v_{LC}^{\downarrow}(P)\right].
\label{midB}
\end{equation}
With
Eqs.~(\ref{midA}) and (\ref{midB}) we can write the expressions given in
Eqs.~(\ref{firststepA}), (\ref{firststepB}) in
an invariant form
\begin{eqnarray}\nonumber
&&\bar{u}_{LC}(k_1)\left(1+\gamma_0\right)\gamma_5 i\gamma_0\gamma_2
\left(1+\gamma_0\right)\bar{u}_{LC}^{\top}(k_2)\\\nonumber
&&=
\frac{1}{M^2_{0}}\bar{u}_{LC}(p_1)\left(\gamma\cdot P+M_0\right)\gamma_5 C
\left(\gamma ^{\top}\cdot P+M_0\right)\bar{u}_{LC}^{\top}(p_2)\\
&&=
{2\over M_0}\bar{u}_{LC}(p_1)\left(\gamma \cdot P+M_0\right)\gamma_ 5 C
\bar{u}_{LC}^{\top}(p_2),
\label{mfa}
\end{eqnarray}
where $C=i\gamma_ 0\gamma_ 2$ is the charge conjugation matrix.

{}From inspection, we recognize that the invariant expression may be
derived from the nonrelativistic one by simply replacing the Pauli
spinors in the matrix elements of Eq.~(\ref{nspin}) via
\begin{equation}
\chi_{LC}\rightarrow
N\,(\gamma \cdot P+M_{0})u_{LC}(k^{+},{\bf k}_{\perp},\lambda ),\,
\quad N=2\sqrt{{m x\over (x M_0+m)^{2}+{\bf k}_{\perp}^{2}}},\
\label{MeDi}
\end{equation}
where $M_0^2$ is the free mass squared of the three-quark system of
Eq.~(\ref{invmass}) and $P$
the free total momentum of the system in Lorentz covariant
Bakamjian-Thomas models~\cite{BT}.

In a last step, we can combine the spin invariants shown in
Eq.~(\ref{nspin}) with the corresponding isospin matrix elements so as
to obtain 1$\leftrightarrow$2 symmetric invariants which are then
symmetrized in the $uds$ basis.  All steps combined yield the
relativistic spin-isospin wave function of the nucleon $\psi _N$ in
the covariant form with $G=i\tau_2 C$,
\begin{eqnarray}
\psi_{N}={\cal N}\sum_{\lambda_i}\left[\left(\bar{u}_{1}(\gamma \cdot P+M_{0})
\gamma _{5}G\bar{u}_{2}^{\top}\right)
\left(\bar{u}_{3}u_{\lambda}\right)+(23)1+(31)2\right]u_1u_2u_3,\
\label{psiN}\end{eqnarray}
where the two factors $(\gamma \cdot P+M_{0})$ in the (12) matrix
element are combined into one factor as in Eq.~(\ref{mfa}) and the
third projection factor in the second matrix element
$\left(\bar{u}_{3}u_{\lambda}\right)$ is eliminated by using the Dirac
equation for the nucleon. In Eq.~(\ref{psiN}), the $\bar u_i$ and
$u_i$ are abbreviations of $\bar u_{LC}(p_i,\lambda _i),
u_{LC}(p_i,\lambda _i)$, and the normalization $\cal N$ that includes
the normalization factors from Eq.~(\ref{MeDi}) among others; it
determines the charge form factor of the proton at zero momentum
transfer.

A complete set of relativistic spin-isospin invariants of the form
Eq.~(\ref{mfa}) is given in Table~\ref{tab:1}. The $\otimes$
symbolizes that each $G_i$ consists of the product of two matrix
elements between Dirac (light-cone) spinors, i.e. is of the form
(12)3. The Dirac spinors for the quarks have been omitted for
simplicity. These spin-isospin states are symmetrized in the $uds$
basis~\cite{Fra}, where quarks are treated as distinguishable; if the
third quark is taken to be the down quark, then the up quarks are
symmetrized explicitly in the spin-flavor wave function. That is why
the isospin operator is chosen so as to make the (12) matrix element
symmetric under $1\leftrightarrow 2$.

Clearly, the nucleon wave function of Eq.~(\ref{psiN}) is the
symmetrized combination $\left(G_2+G_6\right)$. From the construction
it should be clear that the normalized wave function including the
Melosh normalization factors of Eq.~(\ref{MeDi}) coincides with the
corresponding one from the Pauli-Melosh basis. In order to obtain a
nucleon wave function component from the $G_i$ of Table~\ref{tab:1}
the Dirac spinors must be included in each invariant as in
Eq.~(\ref{psiN}) and symmetrized, i.e. adding $(23)1+(31)2$ to $(12)3$. In
order to generate a component corresponding to the Pauli-Melosh basis
the projection onto Pauli spinors via $(\gamma \cdot P+M_0)$ for each
quark is required as well.  For $G_1$, the projection factors from
quark 1 and 2 yield $(\gamma \cdot P+M_0)(-\gamma \cdot P+M_0)=0$.
When the projections are similarly applied to the mixed symmetric
spin-isospin invariants $G_3$ and $G_8$, the same channel wave
function $\psi _{N}$ is generated.  Just like $G_1$, $G_5$ and $G_7$
make no contribution to $\psi_N$ either. Hence, just as the
nonrelativistic nucleon spin-isopin wave function is unique, so is its
relativistic generalization of the Pauli-Melosh
basis~\cite{Azn,Dzi,KW}. This is also the case for other baryon wave
functions. These procedures apply similarly to all relativistic spin-flavor 
wave functions for nucleon resonances that will be discussed in the
following. Let us note, however, that mesonic or sea quark-antiquark
Fock components will lead to new spin-flavor wave functions and configuration 
mixing expands the nucleon basis by N* states with the spin-flavor quantum 
numbers of the nucleon.

\subsection{The $N_{\frac{1}{2}^-}^*(1535)$ spin-isospin states }

We start again from the nonrelativistic states for the $P$-wave given
below.  The notation is $|\left[(s_{12}\Hh)\Hh L\right]JM_J\rangle$,
where $L=1$ in this case.  The corresponding Dirac-Melosh state has
been constructed in~\cite{KW}. The Clebsch-Gordan coefficients in the
nonrelativistic state are written as products of matrix elements as
for the nucleon so that
\begin{eqnarray}\nonumber
|\left[(0\Hh)\Hh1\right]\Hh\lambda\rangle&=&-\sqrt{\frac{1}{8\pi}}
\sum_{m_{1}m_{2}m_{3}}
\left(\chi_{m_{1}}^{\dagger}i
\sigma_{2}\chi_{m_{2}}^{\ast}\right)\left(\chi_{m_{3}}^{\dagger}\vec{\sigma}
\cdot (\vec{k}_1-\vec{k}_2)\;
\chi_{\lambda}\right)\chi_{m_{1}}\chi_{m_{2}}\chi_{m_{3}},
\\
|\left[(1\Hh)\Hh1\right]\Hh\lambda\rangle&=&
\sqrt{\frac{1}{24\pi}}
\sum_{m_{1}m_{2}m_{3}}\left(\chi_{m_{1}}^{\dagger}
\vec\sigma i\sigma_{2}\chi_{m_{2}}^{\ast}\right)\cdot
\left(\chi_{m_{3}}^{\dagger}
\vec\sigma \ \vec{\sigma}\cdot(\vec{k}_1-\vec{k}_2)\;
\chi_{\lambda}\right)\chi_{m_{1}}\chi_{m_{2}}\chi_{m_{3}}.
\label{nrspin}
\end{eqnarray}

These states differ from the nonrelativistic nucleon states only by
the P-wave factor $\vec{\sigma}\cdot(\vec{k}_1-\vec{k}_2)$ so that the
relativistic states are directly obtained from the nucleon states by
the replacement $u_{N}\rightarrow \gamma_{5}\,\gamma \cdot
(\tilde{p}_1-\tilde{p}_2)u_{N^{\ast}}$ in Eq.~(\ref{psiN}) and
Table~\ref{tab:1}.  The four-vector
$\tilde{p}_1-\tilde{p}_2=\left(p_1-p_2\right)-((p_1-p_2)\cdot P/P^2)P$
reduces to $\vec{k}_1-\vec{k}_2$ in the baryon rest frame, as it is
the case in the orbital wave functions of the Pauli-Melosh basis. The
use of $\tilde{p}_1-\tilde{p}_2$ guarantees the orthogonality of
different orbital angular momentum states in the Dirac-Melosh basis.
\par
Again, the spin-isospin states will be symmetrized in the $uds$ basis,
where quarks are treated as distinguishable and light quarks are
symmetrized explicitly in the spin-flavor wave function. Therefore, it
suffices to consider the mixed antisymmetric relative momentum
$p_1-p_2$ of Eq.~(\ref{rholam}) since the mixed symmetric
$p_1+p_2-2p_3$ terms will be automatically generated by symmetrizing.
The resulting relativistic expression for the spin-isospin wave
function is then
\begin{eqnarray}
\psi_{N^{\ast}}={\cal N}\sum_{\lambda _i}\left[\left(\bar{u}_{1}(\gamma \cdot
P+M_0)\gamma_{5}\vec{\tau}G\bar{u}_{2}^{\top}\right)
\left(\bar{u}_{3}\gamma_{5}\,\gamma \cdot (\tilde{p}_1-\tilde{p}_2)\vec{\tau}
u_{\lambda}\right)+(23)1+(31)2\right]u_1u_2u_3,
\label{nstwf}
\end{eqnarray}
for the dominant $N^{\ast}_{\frac{1}{2}^-}$(1535) configuration.  The
$\left(\gamma\cdot P+M_0\right)$ factor from the third quark has been
removed using the commutator $[\gamma \cdot P, \gamma \cdot
(p_1-p_2)]$ and $P\cdot (\tilde{p}_1-\tilde{p}_2)=0$. Again, this wave
function coincides with the corresponding one from the Pauli-Melosh
basis.  The complete set of spin invariants is given in
Table~\ref{tab:2}.

\subsection{$N_{\frac{3}{2}^-}^{\ast}(1520)$ Basis States}
We now consider the $J=\frac{3}{2}^-$ resonance and therefore
introduce Rarita-Schwinger spinors.  In the LS-coupling scheme the
nonrelativistic spin wave function of the
$N_{\frac{3}{2}^-}^{\ast}(1520)$ of spin ${3 \over 2}$ and negative
parity in the baryon rest frame is written again in terms of matrix
elements between Pauli spinors using the Wigner-Eckart theorem
\begin{eqnarray}\nonumber
|\left[(0\Hh)\Hh1\right]\Dh\lambda\rangle
&=&
-\sqrt{\frac{3}{8\pi}}\sum_{m_{1}m_{2}m_{3}}\left(\chi_{m_{1}}^{\dagger}i
\sigma _{2}\chi _{m_{2}}^{\ast}\right)\left(\chi _{m_{3}}^{\dagger}
\left(k_{1\;\nu}-k_{2\;\nu}\right)
u_{\frac{3}{2}\lambda }^{\nu }\right)\chi _{m_{1}}
\chi _{m_{2}}\chi _{m_{3}},\\
|\left[(1\Hh)\Hh1\right]\Dh\lambda\rangle
&=&
\sqrt{\frac{1}{8\pi }}\sum_{m_{1}m_{2}m_{3}}\left(\chi _{m_{1}}^{\dagger}
\vec{\sigma }i\sigma _{2}\chi _{m_{2}}^{\ast}\right)\cdot
\left(\chi _{m_{3}}^{\dagger}
\vec{\sigma }
\left(k_{1\;\nu}-k_{2\;\nu}\right)
u_{\frac{3}{2}\lambda }^{\nu }\right)
\chi _{m_{1}}\chi _{m_{2}}\chi _{m_{3}}.
\end{eqnarray}
Here we have used Rarita-Schwinger spinors to describe the spin
$\frac{3}{2}$ state and introduced $\epsilon^{\mu}$ via
\begin{eqnarray}
\vec{\epsilon}_m\vec{k}=-\epsilon^{\mu }_{m}(\stackrel{\circ}{P})k_{\mu},
\end{eqnarray}
in the rest system of the baryon, where $\stackrel{\circ}{P}^{\mu
  }=(M_0,\vec{0})$, and $\epsilon ^{\mu }_{m}(\stackrel{\circ}{P})$ is
the spin-1-vector~\cite{WAP2}
\begin{equation}
\epsilon ^{\mu }_{m}(\stackrel{\circ}{P}) = (0,{\bf \epsilon }_{m}).
\end{equation}
With Clebsch-Gordan coefficients the $\epsilon _{m}^{\mu }$ and
$N^{\ast}$ Pauli spinor are combined to a Rarita-Schwinger spinor.

In a general frame, the polarization vectors $\epsilon^{\mu }_{\lambda
  }(P)=L^{~\mu}_{\nu}(P)\epsilon ^{\nu}_{\lambda}(\stackrel{\circ}{P})
$ satisfy the orthogonality and normalization conditions
\begin{eqnarray}\nonumber
P\cdot\epsilon _{+}&=&P\cdot\epsilon _{-}=P\cdot\epsilon _{0}=0,
\\\nonumber
\epsilon ^{\dag}_{+}\cdot\epsilon _{+}&=&\epsilon ^{\dag}_{-}\cdot\epsilon _-=
\epsilon ^{\dag}_{0}\cdot\epsilon _{0}=-1,
\\
\epsilon ^{\dag}_{+}\cdot\epsilon _{-}&=&\epsilon ^{\dag}_{+}\cdot\epsilon _0
=\epsilon ^{\dag}_{-}\cdot\epsilon _{0}=0,
\label{pol}
\end{eqnarray}
so that, more explicitly,
\begin{eqnarray}\nonumber
\epsilon ^{\mu}_{+}&=&-{1\over \sqrt{2}}
\left(0,1,i,2P^{R}/P^{+}\right),
\\\nonumber
\epsilon ^{\mu}_{0}&=&{1\over M_0}
\left(P^{+},P^1,P^2,({\bf P}^2_{\perp}-M_0^2)/P^{+}\right),
\\
\epsilon ^{\mu}_{-}&=&{1\over\sqrt{2}}
\left(0,1,-i,2(P^{L})/P^{+}\right).
\label{eps}
\end{eqnarray}
\par
Again due to proper orthogonality we use the formally covariant
expression $\tilde{p}$ for $\vec{k}$ in the nucleon rest frame.  These
spin wave functions can be written as Lorentz covariant expressions
\begin{eqnarray}\nonumber
I_0 &=& \left(\bar{u}_{1}\gamma_{5}\left(\gamma\cdot P+M_0\right)
C\bar{u}_{2}^{\top}\right)
\left(\bar{u}_{3}(\tilde{p}_{1\nu}-\tilde{p}_{2\nu})
u_{\frac{3}{2}\lambda}^{\nu}\right),\\
I_1 &=& \left(\bar{u}_{1}\left(\gamma\cdot P+M_0\right)\gamma^{\mu}C
\bar{u}_{2}^{\top}
\right)
\left(\bar{u}_{3}\gamma _{\mu }\gamma _{5}
(\tilde{p}_{1\nu}-\tilde{p}_{2\nu})u_{\frac{3}{2}\lambda }^{\nu }\right).
\end{eqnarray}
The construction of these basis states closely follows the rules
outlined in the previous sections. In particular, in order to
construct a P-wave state with orbital angular momentum $L=1$, the
Dirac-Melosh states should contain one momentum $\tilde{p}$ factor. In
addition, in the nonrelativistic limit all basis states should either
be linear combinations of $I_{0}$ or $I_{2}$, or vanish.  A set of
basis states is then given by substituting $u\rightarrow
\tilde{p}_{\nu }u_{\frac{3}{2}\lambda }^{\nu }$ in the nucleon basis,
where $u_{\frac{3}{2}\lambda }^{\nu }$ is the Rarita-Schwinger spinor
using the $uds$ basis with $p\equiv p_{\rho }$.  The construction of
further invariants is restricted by the following constraints of the
Rarita-Schwinger spinors
\begin{eqnarray}
\gamma _{\mu }u_{{3\over 2}\lambda }^{\mu }=0,\quad
P_{\mu }u_{{3\over 2}\lambda }^{\mu }=0,
\label{bed}
\end{eqnarray}
that lead to
\begin{eqnarray}
P_{\mu }u_{\frac{3}{2}\lambda}^{\mu }
={1\over 2}\gamma_{\mu }\gamma \cdot Pu_{\frac{3}{2}\lambda}^{\mu }.
\end{eqnarray}
Hence the associated properly symmetrized wave function may be written as
\begin{eqnarray}
\psi_{N^*}={\cal N}\sum_{\lambda_i}\left[\bar u_1(\gamma \cdot P +M_0)
\gamma _5 \vec{\tau } G\bar u_2^{\top} \bar u_3 \vec{\tau }
(\tilde{p}_{1\nu }-\tilde{p}_{2\nu })u_{{3\over 2}\lambda }^{\nu }
+ (23)1 + (31)2\right]u_1u_2u_3.
\end{eqnarray}
The isospin dependence is determined so as to give symmetric
combinations for the $1\leftrightarrow 2$ matrix element. This wave
function can be constructed directly from the invariants in
Table~\ref{tab:3}, as explained for the nucleon case.

\subsection{Basis States for $N^{\ast}_{\frac{5}{2}^-}(1675)$}

The nonrelativistic representation of the state $|{3\over 2}\times
1\rangle^{\frac{5}{2}}$ is given by
\begin{eqnarray}
\left|\left[(1\Hh)\Dh1\right]\Fh\lambda\right\rangle =
\sqrt{\frac{3}{8\pi }}\sum_{m_{1}m_{2}m_{3}}\left(\chi _{m_{1}}^{\dagger}
\sigma _{\mu }i\sigma _{2}\chi _{m_{2}}^{\ast}\right)
\left(\chi _{m_{3}}^{\dagger}
\left(\tilde{p}_{1\;\nu}-\tilde{p}_{2\;\nu}\right)
u_{\frac{5}{2}
\lambda }^{\mu \nu }\right),
\end{eqnarray}
using the formally covariant representation $\vec{k}\rightarrow
\tilde{p}$.  In the $uds$ basis the momentum is chosen to be the mixed
antisymmetric combination
$\left(\tilde{p}_{1\;\nu}-\tilde{p}_{2\;\nu}\right)$.  Invariants
which contain $p_{\nu }u^{\mu \nu }$ are, for example,
\begin{eqnarray}\nonumber
I_0 &=& \bar{u}_1 M_0\gamma _{\mu }G\bar{u}_2^{\top}
\bar{u}_3(\tilde{p}_{1\nu }-\tilde{p}_{2\nu })
u_{{5\over 2}\lambda }^{\mu \nu }+(23)1+(31)2,
\\\nonumber
I_1 &=& \bar{u}_1 M_0\gamma _{\mu }\gamma _{5}G\bar{u}_2^{\top} \bar{u}_3
(\tilde{p}_{1\nu }-\tilde{p}_{2\nu })u_{{5\over 2}\lambda }^{\mu \nu }
+(23)1+(31)2,
\\\nonumber
I_2 &=& \bar{u}_1iM_0\sigma _{\rho \nu }G\bar{u}_2^{\top}
\bar{u}_3 \gamma ^{\rho }(\tilde{p}_{1\mu }-\tilde{p}_{2\mu })
u_{{5\over 2}\lambda }^{\mu \nu }+(23)1+(31)2,
\\
I_3 &=& \bar{u}_1 i\sigma _{\rho \nu }P^{\nu }G\bar{u}_2^{\top}
\bar{u}_3(\tilde{p}_{1\mu }-\tilde{p}_{2\mu })
u_{{5\over 2}\lambda }^{\mu \rho }+(23)1+(31)2.\
\end{eqnarray}
Further invariants are restricted by the Rarita-Schwinger constraint
given in Eq.~(\ref{bed}).
The corresponding wave function obtained from the Melosh rotated NQM state
takes the standard form
\begin{eqnarray}
\psi_{N^*}={\cal N}\sum_{\lambda _i}\left[\bar{u}_1
(\gamma \cdot P +M_0)\gamma _5 G
\bar{u}_2^{\top} \bar{u}_3(\tilde{p}_{1\mu }-\tilde{p}_{2\mu })
(\tilde{p}_{1\nu }-\tilde{p}_{2\nu })u_{{5\over 2}\lambda }^{\mu\nu }\
+ (23)1 + (31)2\right]u_1u_2u_3.
\end{eqnarray}
\par
Summarizing, the advantages of the Dirac-Melosh basis are the ease and
transparency of (e.g. current) matrix element calculations and the
manifest (kinematic) rotational and Lorentz transformation properties
of the wave functions which follow from the use of free light-cone Dirac 
spinors for the quarks and total momentum motion. Amongst its disadvantages
are the need of Fierz transformations if one wants to rewrite the
(23)1 and (31)2 permutations of wave function components in the
canonical 123 quark order.  These are avoided in the Bargmann-Wigner
basis to be discussed in the following Sects.V and VI.

\section{Symmetrized BW Basis}

The general Bargmann-Wigner (BW) basis~\cite{Ca,BW,MKo} of
relativistic three-particle states contains 64 product states of three
free particle light cone spinors $U^{\lambda},V^{\lambda}$ that
satisfy the free Dirac equations
\begin{eqnarray}
\left(\gamma \cdot P-M_0\right)U^{\lambda }(P)&=&0,\\
\left(\gamma \cdot P+M_0\right)V^{\lambda }(P)&=&0,
\label{diracBW}
\end{eqnarray}
where $P^{\mu}$ is total (free particle) baryon momentum and $M_0$ its
mass.

Restricting this basis to definite parity and positive total spin
projection $S_{z}$ obviously leaves the 16 states $B_{1}$ to $B_{16}$
that are shown in Table~\ref{tab:4}.

The BW basis has several advantages. For one, the nonrelativistic
limit is obtained just by deleting the $V$ spinors, and the extreme
relativistic limit, ($P_{z}\rightarrow\infty$), by setting
$U^{\uparrow}=V^{\uparrow}$ and $U^{\downarrow}=-V^{\downarrow}$.

Moreover, product states with particle permutations are readily
expressed in the BW basis, whereas the Dirac-Melosh basis requires
Fierz transformations~\cite{WAP} when particle indices are
interchanged.  Its disadvantage that spins are not well defined is
removed by symmetrizing the product states appropriately, as will be
shown next.  To this end, $U$ and $V$ Dirac spinors are combined to
form the fundamental representation of a $SU(2)_{R}$ so that the
spin-isospin wave function of a baryon is represented as
$$ SU(2)_{S}\otimes SU(2)_{R}\otimes SU(3)_{F},$$
and each state has well defined spin and permutation symmetry in
the first two particles.

For three quarks the representations of $SU_S(2)\otimes
SU_R(2)\subseteq SU(4)$ are displayed in the Table~\ref{tab:5}, where
the functions $\xi$ and $\varphi$ for different permutation symmetries
$(S,A,M_S,M_A)$ are defined in Table~\ref{tab:6} and
Table~\ref{tab:7}.  We denote completely symmetric $SU(2)_{R}\otimes
SU(2)_{S}$ states by $S$, primed for spin ${3 \over 2}$ and unprimed
for spin ${1 \over 2}$, mixed symmetric by $s$, mixed antisymmetric by
$a$ and completely antisymmetric by $A$.

To further simplify the notation we use $\varphi^\prime$ for $m_S=+{3
  \over 2}$ states and $\varphi$ for $m_S=+{1 \over 2}$ states,
$\xi^\prime$ for $m_R=3$ and $\xi$ for $m_R=-1$ positive parity
states.  The functions $S_1^{\prime},...,A$ not shown in Table~\ref{tab:5} are
given in terms of product functions $\xi\varphi$ in Table~\ref{tab:8}.

For the nucleon we get three different basis states to
construct a mixed symmetric basis with $S=\frac{1}{2}\;m_s=\frac{1}{2}$,
\begin{eqnarray}\nonumber
\xi'_S \varphi_{M_S}
&=&|UUU\rangle\times |{1 \over \sqrt{2}}\left(\uparrow\downarrow
-\downarrow\uparrow\right)\uparrow\rangle= {1 \over\sqrt{2}}
\left(U^{\uparrow}U^{\downarrow}-U^{\downarrow}U^{\uparrow}
\right)U^{\uparrow},\\\nonumber
\xi_S \varphi_{M_S}
&=&|{1\over\sqrt{3}}\left(VVU+UVV+VUV\right)\rangle\times|
{1\over\sqrt{2}}\left(\uparrow\downarrow
-\downarrow\uparrow\right)\uparrow\rangle, \\\nonumber
(\xi_{M_S}\varphi_{M_S}-\xi_{M_A}\varphi_{M_A})
&=&\frac{1}{\sqrt{18}}\left[\left(V^{\uparrow}U^{\downarrow}
+U^{\downarrow}V^{\uparrow}\right)V^{\uparrow}
+\left(U^{\uparrow}V^{\uparrow}
+V^{\uparrow}U^{\uparrow}\right)V^{\downarrow}\right.\\
&&\left.+\left(V^{\uparrow}V^{\downarrow}
+V^{\downarrow}V^{\uparrow}\right)U^{\uparrow}\right].
\end{eqnarray}

In the nonrelativistic case the $V$ spinors vanish so that only the
term with $m_R=3$ survives. This way the spin wave function is defined
with proper total spin and proper pair permutation symmetry.  The spin
${1\over 2}$ states~\cite{Ca,MKo} are
\begin{eqnarray}\nonumber
a_{1}&=&\frac{1}{\sqrt{2}}\left(U^{\uparrow}U^{\downarrow}
-U^{\downarrow}U^{\uparrow}\right)U^{\uparrow}
\\\nonumber
a_{2}&=&\frac{1}{\sqrt{6}}\left[\left(V^{\uparrow}V^{\downarrow}
-V^{\downarrow}V^{\uparrow}\right)U^{\uparrow}
+\left(U^{\uparrow}V^{\downarrow}-U^{\downarrow}V^{\uparrow}
+V^{\uparrow}U^{\downarrow}
-V^{\downarrow}U^{\uparrow}\right)V^{\uparrow}\right]
\\\nonumber
a_{3}&=&\frac{1}{\sqrt{6}}\left[-\left(V^{\uparrow}V^{\downarrow}
-V^{\downarrow}V^{\uparrow}\right)U^{\uparrow}
+\left(V^{\uparrow}U^{\downarrow}
-U^{\downarrow}V^{\uparrow}\right)V^{\uparrow}
+\left(U^{\uparrow}V^{\uparrow}
-V^{\uparrow}U^{\uparrow}\right)V^{\downarrow}\right]
\\\nonumber
s_{1}&=&\frac{1}{\sqrt{6}}\left[\left(U^{\uparrow}U^{\downarrow}
+U^{\downarrow}U^{\uparrow}\right)U^{\uparrow}
-2U^{\uparrow}U^{\uparrow}U^{\downarrow}\right]
\\\nonumber
s_{2}&=&\frac{1}{\sqrt{18}}\left[\left(
V^{\uparrow}U^{\downarrow}+V^{\downarrow}U^{\uparrow}
+U^{\uparrow}V^{\downarrow}+U^{\downarrow}V^{\uparrow}\right)V^{\uparrow}
\right]
\\\nonumber
&&+\frac{1}{\sqrt{18}}\left[-2\left(U^{\uparrow}V^{\uparrow}
+V^{\uparrow}U^{\uparrow}\right)V^{\downarrow}
+\left(V^{\uparrow}V^{\downarrow}
+V^{\downarrow}V^{\uparrow}\right)U^{\uparrow}
-2V^{\uparrow}V^{\uparrow}U^{\downarrow}\right]
\\\nonumber
s_{3}&=&\frac{1}{\sqrt{18}}\left[\left(V^{\uparrow}U^{\downarrow}
+U^{\downarrow}V^{\uparrow}\right)V^{\uparrow}
+\left(U^{\uparrow}V^{\uparrow}
+V^{\uparrow}U^{\uparrow}\right)V^{\downarrow}
+\left(V^{\uparrow}V^{\downarrow}
+V^{\downarrow}V^{\uparrow}\right)U^{\uparrow}\right]
\\\nonumber
&&+\frac{1}{\sqrt{18}}\left[-2\left(U^{\uparrow}V^{\downarrow}
+V^{\downarrow}U^{\uparrow}\right)V^{\uparrow}
-2V^{\uparrow}V^{\uparrow}U^{\downarrow}\right]
\\\nonumber
S&=&\frac{1}{\sqrt{18}}\left[\left(
V^{\uparrow}V^{\downarrow}+V^{\downarrow}V^{\uparrow}\right)U^{\uparrow}
+\left(U^{\uparrow}V^{\downarrow}
+V^{\downarrow}U^{\uparrow}\right)V^{\uparrow}
+\left(V^{\uparrow}U^{\uparrow}
+U^{\uparrow}V^{\uparrow}\right)V^{\downarrow}\right]
\\\nonumber
&&+\frac{1}{\sqrt{18}}\left[-2U^{\downarrow}V^{\uparrow}V^{\uparrow}
-2V^{\uparrow}U^{\downarrow}V^{\uparrow}
-2V^{\uparrow}V^{\uparrow}U^{\downarrow}\right]
\\
A&=&\frac{1}{\sqrt{6}}\left[\left(U^{\uparrow}V^{\downarrow}
-V^{\downarrow}U^{\uparrow}\right)V^{\uparrow}
-\left(V^{\uparrow}V^{\downarrow}
-V^{\downarrow}V^{\uparrow}\right)U^{\uparrow}+
\left(V^{\uparrow}U^{\uparrow}
-U^{\uparrow}V^{\uparrow}\right)V^{\downarrow}\right],
\label{nuk}\end{eqnarray}
and the spin ${3\over 2}$ states~\cite{Ca,MKo}
\begin{eqnarray}\nonumber
a'_{1}&=&\frac{1}{\sqrt{6}}\left[-\left(V^{\uparrow}
U^{\uparrow}-U^{\uparrow}V^{\uparrow}\right)V^{\downarrow}+\left(
V^{\uparrow}U^{\downarrow}-U^{\uparrow}V^{\downarrow}\right)V^{\uparrow}+
\left(V^{\downarrow}U^{\uparrow}-U^{\downarrow}V^{\uparrow}\right)
V^{\uparrow}\right]
\\\nonumber
a'_{2}&=&\frac{1}{\sqrt{2}}\left(V^{\uparrow}
U^{\uparrow}-U^{\uparrow}V^{\uparrow}\right)V^{\uparrow}
\\\nonumber
s'_{1}&=&\frac{1}{\sqrt{18}}\left[\left(V^{\uparrow}U^{\uparrow}
+U^{\uparrow}V^{\uparrow}\right)V^{\downarrow}
+\left(V^{\uparrow}U^{\downarrow}
+U^{\uparrow}V^{\downarrow}\right)V^{\uparrow}
+\left(V^{\downarrow}U^{\uparrow}
+U^{\downarrow}V^{\uparrow}\right)V^{\uparrow}\right]
\\\nonumber
&&+\frac{1}{\sqrt{18}}\left[-2V^{\uparrow}V^{\uparrow}U^{\downarrow}
-2V^{\uparrow}V^{\downarrow}U^{\uparrow}
-2V^{\downarrow}V^{\uparrow}U^{\uparrow}\right]
\\\nonumber
s'_{2}&=&\frac{1}{\sqrt{6}}\left[\left(V^{\uparrow}U^{\uparrow}
+U^{\uparrow}V^{\uparrow}\right)V^{\uparrow}
-2V^{\uparrow}V^{\uparrow}U^{\uparrow}\right]
\\\nonumber
S'_{1}&=&\frac{1}{\sqrt{3}}\left[U^{\uparrow}
U^{\uparrow}U^{\downarrow}+U^{\uparrow}U^{\downarrow}U^{\uparrow}
+U^{\downarrow}U^{\uparrow}U^{\uparrow}\right]
\\\nonumber
S''_{1}&=& U^{\uparrow}U^{\uparrow}U^{\uparrow}
\\\nonumber
S'_{2}&=&\frac{1}{3}\left[V^{\uparrow}V^{\uparrow}U^{\downarrow}
+V^{\uparrow}V^{\downarrow}U^{\uparrow}
+V^{\downarrow}V^{\uparrow}U^{\uparrow}+
V^{\uparrow}U^{\uparrow}V^{\downarrow}
+V^{\uparrow}U^{\downarrow}V^{\uparrow}
+V^{\downarrow}U^{\uparrow}V^{\uparrow}\right]
\\\nonumber
&&+\frac{1}{3}\left[U^{\uparrow}V^{\uparrow}V^{\downarrow}
+U^{\uparrow}V^{\downarrow}V^{\uparrow}
+U^{\downarrow}V^{\uparrow}V^{\uparrow}\right]
\\
S''_{2}&=&\frac{1}{\sqrt{3}}\left[V^{\uparrow}
V^{\uparrow}U^{\uparrow}+V^{\uparrow}U^{\uparrow}V^{\uparrow}
+U^{\uparrow}V^{\uparrow}V^{\uparrow}\right].
\label{del}\end{eqnarray}

The orbital function which have $(S,M_A,M_S)$ symmetries is given by
\begin{equation}
{\cal Y}_{Lm_L} = \sum_{m_{\rho} m_{\lambda}}
\braket{\ell_{\rho} m_\rho l_{\lambda} m_{\lambda}}{L m_L}
Y_{\ell_{\rho} m_\rho}(\hat{\bf p}_\rho)
Y_{l_{\lambda}m_{\lambda}}(\hat{\bf p}_\lambda )=
\koppl{Y^{[\ell_{\rho}]}(\hat{\bf p}_\rho)}
{Y^{[\ell_{\lambda}]}(\hat{\bf p}_\lambda )}{[L]}_{m_L}.
\end{equation}

As an example for the $L=1$ case we want to propose a construction
method for BW basis states for the $N_{\frac{1}{2}^-}^*(1535)$
resonance.  As in the previous subsection, we consider only the
antisymmetric orbital function
\begin{eqnarray}
\psi_{M_A}=k_{\rho }Y_{1m}(\hat{{\bf k}}_{\rho })={\cal Y}_{1m}
\left(k_{\rho }\right),
\end{eqnarray}
where $k_{\rho }$ is the relative momentum in the nucleon rest frame
denoted by $k$ in the following for simplicity.  Depending on the
$z$-projection, $\psi _{M_A}$ is a function of the rest frame momentum
variables $k_z$, $k_R$ or $k_L$
\begin{eqnarray}
{\cal Y}_{10}(k)=\sqrt{\frac{3}{4\pi}}k_{z},\quad
{\cal Y}_{1\pm 1}(k)=\mp\sqrt{\frac{3}{8\pi}}\left(k_{x}\pm
ik_{y}\right).
\end{eqnarray}

We generalize these momentum variables to a general frame using
Eq.~(\ref{kin}) in the following manner
\begin{eqnarray}
k_{0} = \frac{P \cdot p}{M_0},\
k_{z} =  -{P \cdot p\over M_0} + {M_0 p^+\over P^+},\
k_{R} = p_{R}-{P_{R}p^+\over P^+},\
k_{L} = p_{L}-{P_{L}p^+\over P^+}.
\end{eqnarray}
Now the orbital function $\psi_{M_A}$ can be coupled with positive
parity antisymmetric spin $\frac{1}{2}$ functions $\chi_{M_A}^{{1\over
    2}+}=a_1,a_2,a_3$ or symmetric functions $\chi_{M_{S}}^{{1\over
    2}+}=s_1,s_2,s_3$ shown in Tables ~\ref{tab:5} and ~\ref{tab:8},
by means of Clebsch-Gordon coefficients
\begin{eqnarray}\nonumber
\left[\chi_{M}^{\frac{1}{2}+}\times{\cal Y}_{1}\right]^{\frac{1}{2}}
 = \sqrt{\frac{2}{3}}\chi_{M,+\frac{1}{2}}^{\frac{1}{2}+}{\cal Y}_{11}-
\sqrt{\frac{1}{3}}\chi_{M,-\frac{1}{2}}^{\frac{1}{2}+}{\cal Y}_{10}\\
 = -\frac{1}{\sqrt{4\pi}}\left[\left(p_R-\frac{P_R p^+}{P^+}\right)
\chi_{M,-\frac{1}{2}}^{\frac{1}{2}+}
+\left(p_L-\frac{P_L p^+}{P^+}\right)\chi_{M,+\frac{1}{2}}^{\frac{1}{2}+}
\right].
\label{l=1BW}
\end{eqnarray}


\section{Transformation of Dirac-Melosh into BW Basis}
We now expand the Dirac-Melosh states $G_i$ of the nucleon basis in
Table~\ref{tab:1} (of Sect.IV) into the Bargmann-Wigner states $B_i$
of Table~\ref{tab:4} (in the previous Sect.V). This will be done first
for the nucleon and subsequently for the other nucleon resonances,
where the abbreviation $G=Ci\tau _2$ is used.

\subsection{The nucleon spin-isospin states}
The second nucleon term $G_2=M_0\gamma _5 Ci\tau_2\otimes U$ in
Eq.~(\ref{psiN}) is an even $\gamma $-matrix coupling $U$ with $U$ and
$V$ with $V$, combined with (12) antisymmetric spin structure.
Therefore,
\begin{eqnarray}
[\gamma _5 C\otimes U^{\uparrow}]_{123} = (UU+VV)U\times
(\uparrow\downarrow-\downarrow\uparrow)\uparrow
=\left(U^{\uparrow}U^{\downarrow}-U^{\downarrow}U^{\uparrow}
+V^{\uparrow}V^{\downarrow}-V^{\downarrow}V^{\uparrow}\right)U^{\uparrow}.
\label{bw2}
\end{eqnarray}
In the other nucleon term $G_6=M_0\gamma _0 \gamma _5 Ci\tau_2\otimes
U$ in the rest frame of the three-quark system in Eq.~(\ref{psiN}),
$\gamma _0$ changes the sign of the $VV$ term so that
\begin{eqnarray}
[\gamma _0 \gamma _5 C\otimes U^{\uparrow}]_{123}=(UU-VV)U\times
(\uparrow\downarrow-\downarrow\uparrow)\uparrow.
\label{bw3}
\end{eqnarray}
In order to check, e.g. the coefficient of the
$V^{\uparrow}V^{\downarrow}$ term in Eq.~(\ref{bw2}) we have to
calculate the overlap matrix element
\begin{eqnarray}
\bar V^{\uparrow}_{\alpha }\bar V^{\downarrow}_{\beta }
(\gamma _5C)_{\alpha \beta }
=-\bar V^{\uparrow}_{\alpha } V^{\uparrow}_{\alpha }
={1\over 4M_0}{\rm Tr}[(1-\gamma _5\gamma \cdot \epsilon _0)
(M_0-\gamma \cdot P)]=1,
\end{eqnarray}
using
$$\bar V^{\downarrow}_{\beta }(\gamma _5C)_{\alpha \beta }
=-V^{\uparrow}_{\alpha },$$
The other terms in Eq.~(\ref{bw2}) can be
checked similarly.  Much the same reasoning yields for $G_1$ and $G_5$
\begin{eqnarray}\nonumber
[C \otimes \gamma_5  U^{\uparrow}]_{123}&=&
(UV+VU)V\times(\uparrow\downarrow-\downarrow\uparrow)\uparrow,
\\
~[\gamma_0 C \otimes \gamma_5  U^{\uparrow}]_{123} &=& (UV-VU)V\times
(\uparrow\downarrow-\downarrow\uparrow)\uparrow.
\label{b1}
\end{eqnarray}
These relations correspond to the spinor identities
\begin{eqnarray}\nonumber
C &=&(UV+VU)\times(\uparrow\downarrow-\downarrow\uparrow)
=U^{\uparrow}V^{\downarrow}-U^{\downarrow}V^{\uparrow}
+V^{\uparrow}U^{\downarrow}-V^{\downarrow}U^{\uparrow},
\\\nonumber
\gamma_5 C &=&(UU+VV)\times(\uparrow\downarrow-\downarrow\uparrow)
=U^{\uparrow}U^{\downarrow}-U^{\downarrow}U^{\uparrow}+
V^{\uparrow}V^{\downarrow}-V^{\downarrow}V^{\uparrow},
\end{eqnarray}
\begin{eqnarray}\nonumber
\left(\gamma \cdot P+M_0\right)\gamma_5 C &=&
2M_0UU\times(\uparrow\downarrow-\downarrow\uparrow) =
2M_0\left(U^{\uparrow}U^{\downarrow}-U^{\downarrow}U^{\uparrow}\right),
\\
\left(M_0-\gamma \cdot P\right)C &=&
M_0(UV+VU)(\uparrow\downarrow-\downarrow\uparrow).
\label{a1}
\end{eqnarray}
The vector term $G_3$ has a more complicated spin-isospin structure~\cite{WAP}
\begin{eqnarray}\nonumber\lefteqn{
[\gamma^{\mu } C \otimes\gamma _5\gamma _{\mu } U^{\uparrow}]_{123}}
\\&&
=(UU-VV)U\times(-2\uparrow\uparrow\downarrow+\uparrow\downarrow\uparrow
+\downarrow\uparrow\uparrow)+(UV-VU)V\times(\uparrow\downarrow
-\downarrow\uparrow)\uparrow,
\end{eqnarray}
etc.  In summary, any Dirac--Melosh basis state of Table~\ref{tab:1}
may be written as a linear combination of Bargmann--Wigner basis
states
\begin{equation}
G_i=\sum_{j=1,...,16}c_jB_j,
\label{gbw}
\end{equation}

using the above (Eq.~(\ref{a1})) and the following spinor identities

\begin{eqnarray}\nonumber
U^{\uparrow}\bar U^{\uparrow}&=&{1\over 4M_0}(1+\gamma _5\gamma \cdot
\epsilon _0)(M_0+\gamma \cdot P)
\\\nonumber
U^{\downarrow}\bar U^{\downarrow}&=&{1\over 4M_0}(1-\gamma _5\gamma \cdot
\epsilon _0)(M_0+\gamma \cdot P)
\\\nonumber
U^{\uparrow}\bar U^{\downarrow}&=&{\sqrt{2}\over 4M_0}\gamma _5\gamma \cdot
\epsilon _{+}(M_0+\gamma \cdot P)
\\\nonumber
U^{\downarrow}\bar U^{\uparrow}&=&-{\sqrt{2}\over 4M_0}\gamma _5\gamma \cdot
\epsilon _{-}(M_0+\gamma \cdot P)
\\\nonumber
U^{\uparrow}\bar V^{\uparrow}&=&-{1\over 4M_0}(\gamma _5-\gamma \cdot
\epsilon _0)(M_0-\gamma \cdot P)
\\\nonumber
U^{\downarrow}\bar V^{\downarrow}&=&-{1\over 4M_0}(\gamma _5+\gamma \cdot
\epsilon _0)(M_0-\gamma \cdot P)
\\\nonumber
U^{\uparrow}\bar V^{\downarrow}&=&{\sqrt{2}\over 4M_0}\gamma \cdot
\epsilon _{+}(M_0-\gamma \cdot P)
\\\nonumber
U^{\downarrow}\bar V^{\uparrow}&=&-{\sqrt{2}\over 4M_0}\gamma \cdot
\epsilon _{-}(M_0-\gamma \cdot P)
\\\nonumber
V^{\uparrow}\bar U^{\uparrow}&=&{1\over 4M_0}(\gamma _5+\gamma \cdot
\epsilon _0)(M_0+\gamma \cdot P)
\\\nonumber
V^{\downarrow}\bar U^{\downarrow}&=&{1\over 4M_0}(\gamma _5-\gamma \cdot
\epsilon _0)(M_0+\gamma \cdot P)
\\\nonumber
V^{\uparrow}\bar U^{\downarrow}&=&{\sqrt{2}\over 4M_0}\gamma \cdot
\epsilon _{+}(M_0+\gamma \cdot P)
\\\nonumber
V^{\downarrow}\bar U^{\uparrow}&=&-{\sqrt{2}\over 4M_0}\gamma \cdot
\epsilon _{-}(M_0+\gamma \cdot P)
\\\nonumber
V^{\uparrow}\bar V^{\uparrow}&=&-{1\over 4M_0}(1-\gamma _5\gamma \cdot
\epsilon _0)(M_0-\gamma \cdot P)
\\\nonumber
V^{\downarrow}\bar V^{\downarrow}&=&-{1\over 4M_0}(1+\gamma _5\gamma \cdot
\epsilon _0)(M_0-\gamma \cdot P)
\\\nonumber
V^{\uparrow}\bar V^{\downarrow}&=&{\sqrt{2}\over 4M_0}\gamma _5\gamma \cdot
\epsilon _{+}(M_0-\gamma \cdot P)
\\
V^{\downarrow}\bar V^{\uparrow}&=&-{\sqrt{2}\over 4M_0}\gamma _5\gamma \cdot
\epsilon _{-}(M_0-\gamma \cdot P),
\label{proj}
\end{eqnarray}
where the polarization vectors~\cite{WAP2} $\epsilon _{\lambda }^{\mu }$
were given in Eq.~(\ref{eps}).
\par
It is instructive to note that the Melosh rotation can also be written as
$\bar u_{\lambda _i}(p_i)U^{\lambda }(P)$ since
\begin{eqnarray}\nonumber
\bar u^{\uparrow}(p_i)U^{\uparrow}&=&(p_{i}^{+} M_0+m_i P^+)/
(2\sqrt{m_iM_0p_{i}^{+}P^+}),
\\\nonumber
\bar u^{\uparrow}(p_i)U^{\downarrow}
&=&(P^{+}p_i^{L}-P^Lp_i^+)/(2\sqrt{m_iM_0p_{i}^{+}P^+}),
\\
\bar u^{\downarrow}(p_i)U^{\uparrow}
&=&(P^Rp_i^{+}-P^{+}p_{i}^{R})/(2\sqrt{m_iM_0p_{i}^{+}P^+}).
\label{komel}
\end{eqnarray}
\par
In order to illustrate the usefulness of the BW basis let us first
consider the nucleon in the $uds$ basis where quarks are treated as
distinguishable. For example, if the $d$ quark in the proton is taken
as the third quark, then only the two up quarks are symmetrized in the
spin-flavor wave function.  Rewriting the proton wave function with
the spin up mixed antisymmetric function $\chi_{M_A}^{\uparrow}$ and
the mixed antisymmetric isospin part $\phi_{M_A}$ it becomes
\begin{eqnarray}\nonumber
2\phi_{M_A}\chi_{M_A}^{\uparrow}
&=&(ud-du)u\left[(\uparrow\downarrow-\downarrow\uparrow)\uparrow
\times UUU\right]
\\\nonumber
&=&u_2d_3u_1\left[(\uparrow_2\downarrow_3-\downarrow_2\uparrow_3)\uparrow_1
\times UUU\right]
-d_3u_1u_2\left[(\uparrow_3\downarrow_1-\downarrow_3\uparrow_1)\uparrow_2
\times UUU\right]
\\\nonumber
&=&u_1u_2d_3\left[(\uparrow_1\uparrow_2\downarrow_3-
\uparrow_1\downarrow_2\uparrow_3
-\downarrow_1\uparrow_2\uparrow_3+\uparrow_1\uparrow_2\downarrow_3)
\times UUU\right]
\\
&=&uud\left[(2\uparrow\uparrow\downarrow-\uparrow\downarrow\uparrow
-\downarrow\uparrow\uparrow)
\times UUU\right]=-\sqrt{6}uud\chi_{M_S}^{\uparrow},
\end{eqnarray}
which obviously leads to a mixed symmetric component.
Writing the relativistic proton wave function of Eq.~(\ref{psiN}) in the same
123 quark order of the BW basis starting from
\begin{eqnarray}\nonumber
[G_2+G_6]_{123}=\left(U^{\uparrow}U^{\downarrow}-U^{\downarrow}
U^{\uparrow}\right)U^{\uparrow},
\end{eqnarray}
now easily yields
\begin{eqnarray}
[G_2+G_6]_{231}=[\left(U^{\uparrow}U^{\downarrow}-U^{\downarrow}U^{\uparrow}
\right)U^{\uparrow}]_{231}
=U^{\uparrow}U^{\uparrow}U^{\downarrow}-
U^{\uparrow}U^{\downarrow}U^{\uparrow},
\end{eqnarray}
and hence
\begin{eqnarray}\nonumber
[G_2+G_6]_{231}-[G_2+G_6]_{312}
&=&U^{\uparrow}U^{\uparrow}U^{\downarrow}-
U^{\uparrow}U^{\downarrow}U^{\uparrow}
-U^{\downarrow}U^{\uparrow}U^{\uparrow}+U^{\uparrow}U^{\uparrow}U^{\downarrow}
\\
&=&2U^{\uparrow}U^{\uparrow}U^{\downarrow}-
U^{\uparrow}U^{\downarrow}U^{\uparrow}
-U^{\downarrow}U^{\uparrow}U^{\uparrow},
\end{eqnarray}
the same result. The corresponding lines in the Dirac-Melosh basis
involve the much more complicated Fierz transformations~\cite{WAP} to
rewrite the (23)1 and (31)2 forms in the canonical 123 order of spin
invariants, which are avoided here in the BW basis, which is a
definite advantage.

\subsection{The $N_{\frac{1}{2}^-}^*(1535)$ spin-isospin states }
Now we consider the $N^{\ast}_{\frac{1}{2}^-}(1535)$
states of negative parity in the
Dirac-Melosh basis, shown in Table~\ref{tab:2} and relate them to the
symmetrized BW basis states in Eq.~(\ref{b1}) and Eq.~(\ref{a1}). To
that end we expand the $\gamma \cdot (p_1-p_2)$ part, and here
abbreviating $p_1-p_2=p_{\rho }=p$,
\begin{eqnarray}\nonumber
\gamma \cdot p\; U^{\uparrow} & = & a_{p} U^{\uparrow}
+ c_{p} V^{\uparrow} +d_{p} V^{\downarrow}
\\\nonumber
\gamma \cdot p\; U^{\downarrow} & = & a_{p} U^{\downarrow} +d_{p}^*
V^{\uparrow}- c_{p}V^{\downarrow}
\\\nonumber
\gamma \cdot p\; V^{\uparrow} & = &  -c_{p} U^{\uparrow}
- d_{p} U^{\downarrow}- a_{p} V^{\uparrow}
\\
\gamma \cdot p\; V^{\downarrow} & = & -d_{p}^* U^{\uparrow} + c_{p}
U^{\downarrow} - a_{p} V^{\downarrow},
\label{a0}\end{eqnarray}
with the coefficients
\begin{eqnarray}\nonumber
a_{p} & = & \frac{P \cdot p}{M_0} = k_0
\\\nonumber
c_{p} & = & \frac{M_0 \, p^+}{P^+} -\frac{P \cdot p}{M_0}  = k_z
\\\nonumber
d_{p} & = & p^{R}-\frac{P^{R}p^{+}}{P^{+}} = k_R
\\
d_{p}^{\ast} & = & p^{L}-\frac{P^{L}p^{+}}{P^{+}} = k_L,
\end{eqnarray}
where $p_L=p_x-ip_y, p_R=p_x+ip_y$ as defined before. Therefore the
expanded states of Eq.~(\ref{a0}) to the usual form without $\gamma
\cdot p$ dependence can be straightforward mapped to the BW basis.

As an example we expand the first term of the wave function
$\psi_{N^*}$ given in Eq.~(\ref{nstwf}) in the symmetrized BW basis.
Using
\begin{eqnarray}\left(\gamma \cdot P\right)\left(\gamma \cdot p\right)
=P\cdot p-iP^{\mu }p^{\nu }
\sigma _{\mu \nu }=2P\cdot p-\left(\gamma \cdot p\right)
\left(\gamma \cdot P\right),
\end{eqnarray}
we can write (again with $p=p_1-p_2$)
\begin{eqnarray}\nonumber
\lefteqn{\left[\left(\gamma \cdot P+M_0\right)\gamma _{5}C\right]\otimes
\left[\left(\gamma \cdot P+M_0\right)\gamma \cdot p \gamma _{5}
U^{\uparrow}\right]}\\\nonumber
&&=
\left[\left(\gamma \cdot P+M_0\right)\gamma _{5}C\right]\otimes
\left[2M_0\gamma \cdot p
\gamma _{5}U^{\uparrow}\right]
+\left[\left(\gamma \cdot P+M_0\right)\gamma _{5}C\right]\otimes
\left[2P\cdot p \gamma _{5}U^{\uparrow}\right].
\end{eqnarray}
Now we transform these terms to the symmetrized BW basis.  We then get
with the expression for $\left(\gamma \cdot P+M_0\right)\gamma _{5}C$
given in Eq.~(\ref{a1}) and for $\gamma \cdot p \gamma
_{5}U^{\uparrow}$ in Eq.~(\ref{a0})
\begin{eqnarray}\nonumber
\lefteqn{\left[\left(\gamma \cdot P+M_0\right)\gamma _{5}C\right]\otimes
\left[\left(\gamma \cdot P+M_0\right)\gamma \cdot p\;
\gamma _{5}
U^{\uparrow}\right]}\\\nonumber
&=&
-2M_0\left[\left(\gamma \cdot P+M_0\right)\gamma_{5}C\right]\otimes
\left[\left(\frac{M_0 \, p^+}{P^+} -\frac{P \cdot p}{M_0}\right)U^{\uparrow}
+ \left(p^{R}-\frac{P^{R}p^{+}}{P^{+}}\right)U^{\downarrow}+
\frac{P \cdot p}{M_0}V^{\uparrow}
\right]\\\nonumber
&&+\left[\left(\gamma \cdot P+M_0\right)\gamma _{5}C\right]\otimes
\left[2P\cdot p \; V^{\uparrow}\right]\\\nonumber
&=&-4M_0^2\left[\left(\frac{M_0 \, p^+}{P^+} -\frac{P \cdot p}{M_0}\right)
\left(U^{\uparrow}U^{\downarrow}-U^{\downarrow}U^{\uparrow}\right)U^{\uparrow}
\right.\\
&&+ \left(p^{R}-\frac{P^{R}p^{+}}{P^{+}}\right)
\left(U^{\uparrow}U^{\downarrow}-U^{\downarrow}U^{\uparrow}\right)
U^{\downarrow}\left. \right].
\end{eqnarray}
We find that the wave function $\psi_{N^*}$ given in Eq.~(\ref{nstwf}) can
be expressed in BW basis states shown in Eq.~(\ref{l=1BW}) with
the particular choice $\chi=a_1$.

\section{Conclusion}
We have reviewed several ways of constructing basis states for three
valance quarks to describe baryons on the light cone. These are the
Pauli-Melosh, the Dirac-Melosh, the Bargmann-Wigner and the
symmetrized Bargmann-Wigner bases. All of these bases ensure that wave
functions of moving frames are connected by purely kinematic boosts.
We have compared these bases to each other and discussed their
respective advantages.

The {\em Pauli-Melosh} basis is a minimal extension of the
nonrelativistic basis to ensure proper kinematical boosts that connect
moving frames. Due to the use of Pauli spinors that are properly
Melosh rotated the nonrelativistic coupling scheme can be kept.
Therefore incorporation of angular momentum is possible within the
same algebra on the expense of manifest covariance.

{}From the construction of states in the {\em Dirac-Melosh} basis it is
obvious that these basis states are relativistic generalizations of
the Pauli-Melosh basis states as they ensure the full Dirac spinor
structure.  The {\em Dirac-Melosh} basis can be systematically
enlarged to describe nucleon resonances with non-zero orbital angular
momentum. As an example, the case for $\ell=1$ basis states for
$N^*(1535)$ and $N^*(1520)$ are shown.  The Lorentz covariance of the
Dirac-Melosh basis is manifest.  However, the orthogonality of the
basis states must be proved explicitly.

For the nucleon we have shown that the Dirac-Melosh basis can be
mapped onto the {\em symmetrized Bargmann-Wigner} basis. Orthogonality
and completeness of this basis are fulfilled by construction. The
Melosh rotation is implicitly present through products $\bar
u_{\lambda_i}(p_{i})U^{\lambda}(P)$. The simple symmetrization
procedure leads to advantages in practical calculations.  We extended
the {\em symmetrized Bargmann-Wigner} basis to states with angular
momentum $\ell=1$ and showed that the resulting basis states are
equivalent to the Dirac-Melosh basis states.

As the Pauli-Melosh basis is using two component spinors, the
extension of nonrelativistic codes using standard angular momentum
decomposition should be straightforward at the expense mentioned
previously in Section~\ref{Melsub}.

Many processes, in particular those involving transition amplitudes,
require a covariant treatment of baryons. Even the calculation of
elastic amplitudes involve moving frames (e.g., the Breit frame) to
compare with experimental data, whereas wave functions are usually
given in their respective rest frames. Therefore a manifestly
covariant wave function is clearly appealing in cases were not only
the wave functions in the rest frame of the three quark system is
involved.

Also, the covariant bases potentially allow connecting wave functions
given in the rest system to high energy physics phenomena and,
therefore, investigating processes involved in the interesting
transition regime from the broken chiral symmetry region to the gluon
dominated region of quantum chromodynamics (accessible, e.g., at
TJNAF).

\section{Acknowledgments}
We acknowledge useful discussions with S.\ M.\ Dorkin, V.\ Karmanov
and F.\ Lev.  Part of the work has been done during a Research
Workshop June/July 1997 hosted by the Bogolyubov Laboratory of
Theoretical Physics, JINR, Dubna.

\begin{table}
\caption{\label{tab:1} Relativistic spin invariants for the nucleon $N$,
$|(s_{12}\Hh);\Hh^+\lambda\rangle $.}
\[
\begin{array}{lrcl}\hline\hline
G_1:& M_01G&\otimes &\gamma _{5}u_{\lambda }\\
G_2:& M_0\gamma _{5}G&\otimes &u_{\lambda }\\
G_3:& M_0\gamma ^{\mu }\vec{\tau }G&\otimes &\gamma _{5}\gamma _{\mu }
\vec{\tau }u_{\lambda }\\
G_4:& M_0\gamma ^{\mu }\gamma _{5}G&\otimes& \gamma _{\mu }u_{\lambda }\\
G_5:& \gamma \cdot P\vec{\tau }G&\otimes&\gamma _{5}\vec{\tau }u_{\lambda }\\
G_6:& \gamma \cdot P\gamma _{5}G&\otimes &u_{\lambda }\\
G_7:& M_0\sigma ^{\mu \nu }\vec{\tau }G&\otimes
&\gamma _{5}\sigma _{\mu \nu }\vec{\tau }u_{\lambda }\\
G_8:& i\sigma ^{\mu \nu }P_{\nu }\vec{\tau }G
&\otimes &\gamma _{5}\gamma _{\mu }
\vec{\tau }u_{\lambda }\\\hline\hline
\end{array}
\]
\end{table}
\begin{table}
\caption{\label{tab:2} Relativistic spin invariants for the nucleon resonance
$N^*(1535)$,
$|\left[(s_{12}\Hh)\Hh1\right];\Hh^-\lambda\rangle$.}
\[
\begin{array}{lrcl}\hline\hline
1.& M_01\vec{\tau }G&\otimes & \vec{\tau }\gamma \cdot
(\tilde{p}_1-\tilde{p}_2)u_{\lambda }\\
2.& M_0\gamma _{5}\vec{\tau }G&\otimes &\vec{\tau }
\gamma _{5}\gamma \cdot (\tilde{p}_1-\tilde{p}_2)u_{\lambda }\\
3.& M_0\gamma ^{\mu }G&\otimes &\gamma _{\mu }
\gamma \cdot (\tilde{p}_1-\tilde{p}_2)u_{\lambda }\\
4.& M_0\gamma ^{\mu}\gamma _{5}\vec{\tau }G&\otimes& \vec{\tau }\gamma _{\mu }
\gamma _{5}\gamma \cdot (\tilde{p}_1-\tilde{p}_2)u_{\lambda }\\
5.& \gamma \cdot P G & \otimes
&\gamma \cdot (\tilde{p}_1-\tilde{p}_2)u_{\lambda }\\
6.& \gamma \cdot P\gamma _{5}\vec{\tau }G&\otimes &\vec{\tau }\gamma _{5}
\gamma \cdot (\tilde{p}_1-\tilde{p}_2)u_{\lambda }\\
7.& M_0\sigma ^{\mu \nu }G&\otimes
&\sigma _{\mu \nu }\gamma \cdot (\tilde{p}_1-\tilde{p}_2)u_{\lambda }\\
8.& i\sigma ^{\mu \nu }P_{\nu }G&\otimes &\gamma _{\mu }
\gamma \cdot (\tilde{p}_1-\tilde{p}_2)u_{\lambda }\\ \hline\hline
\end{array}
\]
\end{table}
\begin{table}
\caption{\label{tab:3} Basis states for nucleon resonance $N^{\ast}(1520)$,
$|\left[(s_{12}\Hh)\Hh1\right];\Dh^-\lambda\rangle$.}
\[
\begin{array}{lrcl}\hline\hline
1.& M_01\vec{\tau }G&\otimes &\gamma _{5}\vec{\tau }
(\tilde{p}_1-\tilde{p}_2)_{\nu}u_{\frac{3}{2}\lambda }^{\nu }\\
2.& M_0\gamma _{5}\vec{\tau }G&\otimes &\vec{\tau }
(\tilde{p}_1-\tilde{p}_2)_{\nu }u_{\frac{3}{2}\lambda }^{\nu }\\
3.& M_0\gamma ^{\mu }G&\otimes &\gamma _{5}\gamma _{\mu }
(\tilde{p}_1-\tilde{p}_2)_{\nu }u_{\frac{3}{2}\lambda }^{\nu }\\
4.& M_0\gamma ^{\mu }\gamma _{5}\vec{\tau }G&\otimes& \vec{\tau }\gamma _{\mu }
(\tilde{p}_1-\tilde{p}_2)_{\nu }u_{\frac{3}{2}\lambda }^{\nu }\\
5.&\gamma \cdot P\vec{\tau }G&\otimes&\vec{\tau }\gamma _{5}
(\tilde{p}_1-\tilde{p}_2)_{\nu }u_{\frac{3}{2}\lambda }^{\nu }\\
6.& \gamma \cdot P\gamma _{5}\vec{\tau }G&\otimes
&\vec{\tau }(\tilde{p}_1-\tilde{p}_2)_{\nu }u_{\frac{3}{2} \lambda }^{\nu }\\
7.& M_0\sigma ^{\lambda \rho }G&\otimes
&\gamma _{5}\sigma _{\lambda \rho }(\tilde{p}_1-\tilde{p}_2)_{\nu }
u_{\frac{3}{2}\lambda }^{\nu }\\
8.& i\sigma ^{\lambda \rho }P_{\rho }G&\otimes &\gamma _{5}\gamma _{\lambda }
(\tilde{p}_1-\tilde{p}_2)_{\nu }u_{\frac{3}{2}\lambda }^{\nu }\\ \hline\hline
\end{array}
\]
\end{table}
\begin{table}
\caption{\label{tab:4} BW-basis for positive parity and $S_z>0$.}
\[
\begin{array}{ll}
\hline\hline
B_{1}=U^{\uparrow}U^{\uparrow}U^{\uparrow} &
B_{9}=V^{\uparrow}U^{\uparrow}V^{\uparrow}\\
B_{2}=U^{\uparrow}U^{\uparrow}U^{\downarrow}&
B_{10}=V^{\uparrow}U^{\uparrow}V^{\downarrow}\\
B_{3}=U^{\uparrow}U^{\downarrow}U^{\uparrow}&
B_{11}=V^{\uparrow}U^{\downarrow}V^{\uparrow}\\
B_{4}=U^{\downarrow}U^{\uparrow}U^{\uparrow}&
B_{12}=V^{\downarrow}U^{\uparrow}V^{\uparrow}\\
B_{5}=U^{\uparrow}V^{\uparrow}V^{\uparrow}&
B_{13}=V^{\uparrow}V^{\uparrow}U^{\uparrow}\\
B_{6}=U^{\uparrow}V^{\uparrow}V^{\downarrow}&
B_{14}=V^{\uparrow}V^{\uparrow}U^{\downarrow}\\
B_{7}=U^{\uparrow}V^{\downarrow}V^{\uparrow}&
B_{15}=V^{\uparrow}V^{\downarrow}U^{\uparrow}\\
B_{8}=U^{\downarrow}V^{\uparrow}V^{\uparrow}&
B_{16}=V^{\downarrow}V^{\uparrow}U^{\uparrow}\\
\hline\hline
\end{array}
\]
\end{table}
\begin{table}
\caption{\label{tab:5} Symmetrized BW states for positive parity and $m_j>0$.}
\[
\begin{array}{ccrcl}
\hline\hline
\mbox{dim}[SU(4)]&\mbox{symmetrized BW-states}&\chi^{RS}_{m_R m_S}
\\
\hline
20_S&\xi_S \varphi_S&S'_1,S'_2,S''_1,S''_2\\
&{1\over \sqrt{2}}(\xi_{M_S}\varphi_{M_S}+\xi_{M_A}
\varphi_{M_A})&=S\\[1ex]
20_{M_S}&\xi_S \varphi_{M_S}&s_1,s_2\\
&\xi_{M_S} \varphi_S&s'_1,s'_2\\
&-{1\over \sqrt{2}} (\xi_{M_S}\varphi_{M_S}-\xi_{M_A}
\varphi_{M_A})&=s_3\\[1ex]
20_{M_A}&\xi_S \varphi_{M_A}&a_1,a_2\\
&\xi_{M_A} \varphi_S&a'_1,a'_2\\
&{1\over \sqrt{2}} (\xi_{M_S}\varphi_{M_A}
+\xi_{M_A}\varphi_{M_S})&=a_3\\[1ex]
4_A&{1\over \sqrt{2}} (\xi_{M_S}\varphi_{M_A}-\xi_{M_A}
\varphi_{M_S})&=A\\
\hline\hline
\end{array}
\]
\end{table}
\begin{table}
\caption{\label{tab:6} $SU(2)$ spin states for three quarks.}
\[
\begin{array}{ccccc}
\hline\hline
m_S&+\frac{3}{2}&+\frac{1}{2}&-\frac{1}{2}&-\frac{3}{2}\\
\hline
\varphi_S&\ket{\uparrow\uparrow\uparrow}
&\frac{1}{\sqrt{3}}\ket{\uparrow\downarrow\uparrow
+\downarrow\uparrow\uparrow+\uparrow\uparrow\downarrow}
&\frac{1}{\sqrt{3}}\ket{\uparrow\downarrow\downarrow
+\downarrow\uparrow\downarrow+\downarrow\downarrow\uparrow}
&\ket{\downarrow\downarrow\downarrow}\\
\varphi_{M_S}&&\frac{1}{\sqrt{6}}\ket{\uparrow\downarrow\uparrow
+\downarrow\uparrow\uparrow-2\uparrow\uparrow\downarrow}
&-\frac{1}{\sqrt{6}}\ket{\uparrow\downarrow\downarrow
+\downarrow\uparrow\downarrow-2\downarrow\downarrow\uparrow}&\\
\varphi_{M_A}&&{1\over \sqrt{2}}\ket{\uparrow\downarrow\uparrow
-\downarrow\uparrow\uparrow}&-{1\over \sqrt{2}}\ket{\downarrow\uparrow
\downarrow-\uparrow\downarrow\downarrow}&\\
\hline\hline
\end{array}
\]
\end{table}
\begin{table}
\caption{\label{tab:7} $SU(2)$ $R$-spin states for three quarks. The
respective parity is given by $\pi$.}
\[
\begin{array}{ccccc}
\hline\hline
m_R&+3&+1&-1&-3\\
\pi&+&-&+&-\\
\hline
\xi_S&\ket{UUU}&\frac{1}{\sqrt{3}}\ket{UVU+VUU+UUV}
&\frac{1}{\sqrt{3}}\ket{UVV+VUV+VVU}&\ket{VVV}\\
\xi_{M_S}&&\frac{1}{\sqrt{6}}\ket{UVU+VUU-2UUV}
&-\frac{1}{\sqrt{6}}\ket{UVV+VUV-2VVU}&\\
\xi_{M_A}&&{1\over \sqrt{2}}\ket{UVU-VUU}&-{1\over \sqrt{2}}\ket{VUV-UVV}&\\
\hline\hline
\end{array}
\]
\end{table}
\begin{table}
\caption{\label{tab:8} Explicit form of the functions $S^{\prime}\dots
  A$ not given in  Table V.}
\[
\begin{array}{llll}
\hline\hline
S'_1=\xi'_S\varphi_S&S'_2=\xi_S\varphi_S
&S''_1=\xi'_S\varphi'_S&S''_2=\xi_S\varphi'_S\\
s_1=\xi'_S \varphi_{M_S}&s_2=\xi_S \varphi_{M_S}&&\\
s'_1=\xi_{M_S} \varphi_S&s'_2=\xi_{M_S} \varphi'_S&&\\
a_1=\xi'_S \varphi_{M_A}&a_2=\xi_S \varphi_{M_A}&&\\
a'_1=\xi_{M_A} \varphi_S&a'_2=\xi_{M_A} \varphi'_S&&\\
\hline\hline
\end{array}
\]
\end{table}

\end{document}